\documentclass{aa}
\usepackage{graphicx}
\usepackage{bm,natbib,url,wasysym}
\usepackage[varg]{txfonts}
\usepackage{xcolor}
\usepackage{amsmath}
\bibpunct{(}{)}{;}{a}{}{,} 
\graphicspath{{./fig/}{./png/}}

\newcommand{\brac}[1]{\langle #1 \rangle}
\newcommand{\EQ}{\begin{equation}}
\newcommand{\EN}{\end{equation}}
\newcommand{\EQA}{\begin{eqnarray}}
\newcommand{\ENA}{\end{eqnarray}}

\newcommand{\eq}[1]{Eq.~(\ref{#1})}

\newcommand{\App}[1]{Appendix~\ref{#1}}
\newcommand{\Sec}[1]{Section~\ref{#1}}

\newcommand{\Secs}[2]{Sections~\ref{#1} and \ref{#2}}
\newcommand{\Secss}[2]{Sections~\ref{#1}--\ref{#2}}
\newcommand{\Fig}[1]{Fig.~\ref{#1}}

\newcommand{\Figs}[2]{Figs.~\ref{#1} and \ref{#2}}

\newcommand{\Tab}[1]{Table~\ref{#1}}

\newcommand{\bra}[1]{\langle #1\rangle}

\newcommand{\fluc}[1]{#1^\prime}
\newcommand{\vari}[1]{#1^{\rm v}}
\newcommand{\mean}[1]{\overline#1}
\newcommand{\meanrho}{\overline{\rho}}

{}
{}
{}

{}
{}
{}
{}
{}
{}
{}
{}
{}
\newcommand{\meanBB}{\overline{\mbox{\boldmath $B$}}{}}{}
{}
{}
{}
{}
{}
{}
{}
\newcommand{\meanJJ}{\overline{\mbox{\boldmath $J$}}{}}{}
\newcommand{\meanUU}{\overline{\bm{U}}}

{}
{}
{}

\newcommand{\meanB}{\overline{B}}

\newcommand{\meanU}{\overline{U}}

\newcommand{\flucbb}{\mbox{\boldmath $b^\prime$}{}}{}
\newcommand{\flucuu}{\mbox{\boldmath $u^\prime$}{}}{}
{}
{}

{}

{}
\newcommand{\emf}{{\cal E}}{}

\newcommand{\alphaK}{\alpha_{\rm K}}
\newcommand{\alphaM}{\alpha_{\rm M}}
\newcommand{\alpharr}{\alpha_{rr}}
\newcommand{\alphatt}{\alpha_{\theta\theta}}
\newcommand{\alphapp}{\alpha_{\phi\phi}}
\newcommand{\meanBBp}{\meanBB_{\rm pol}}
\newcommand{\meanUUt}{\meanUU_{\rm tor}}
\newcommand{\meanUUp}{\meanUU_{\rm pol}}
\newcommand{\meanBBt}{\meanBB_{\rm tor}}

\newcommand{\Ot}{\tilde{\Omega}}

\newcommand{\eee}{\hat{\mbox{\boldmath $e$}} {}}

\newcommand{\VV}{\bm{V}}

\newcommand{\uu}{\mbox{\boldmath $u$} {}}

\newcommand{\ssss}{\mbox{\boldmath $\xi$} {}}
\newcommand{\bb}{\mbox{\boldmath $b$} {}}
\newcommand{\BB}{\mbox{\boldmath $B$} {}}

\newcommand{\jj}{\mbox{\boldmath $j$} {}}
\newcommand{\JJ}{\mbox{\boldmath $J$} {}}

\newcommand{\AAA}{\mbox{\boldmath $A$} {}}
\newcommand{\aaaa}{\mbox{\boldmath $a$} {}}
\newcommand{\ee}{\mbox{\boldmath $e$} {}}

\newcommand{\nab}{\mbox{\boldmath $\nabla$} {}}
\newcommand{\OO}{\bm{\Omega}}
\newcommand{\oo}{\mbox{\boldmath $\omega$} {}}

\newcommand{\ddelta}{\mbox{\boldmath $\delta$} {}}
\newcommand{\ggamma}{\mbox{\boldmath $\gamma$} {}}
\newcommand{\aalpha}{\mbox{\boldmath $\alpha$} {}}
\newcommand{\bbeta}{\mbox{\boldmath $\beta$} {}}
\newcommand{\kkappa}{\mbox{\boldmath $\kappa$} {}}

\newcommand{\EMF}{\mbox{\boldmath ${\cal E}$} {}}

\newcommand{\sgn}{{\rm sgn}  \, {}}

\newcommand{\const}{{\rm const}  {}}

\def\Ta{\mbox{\rm Ta}}
\def\Ra{\mbox{\rm Ra}}
\def\Ma{\mbox{\rm Ma}}
\def\Co{\mbox{\rm Co}}

\def\PrSGS{\mbox{\rm Pr}_{\rm SGS}}
\def\Pr{\mbox{\rm Pr}}

\def\Pm{\mbox{\rm Pr}_{\rm M}}
\def\Rm{\mbox{\rm Re}_{\rm M}}

\def\Rey{\mbox{\rm Re}}

\def\Co{\mbox{\rm Co}}

\def\cs{c_{\rm s}}

\def\kf{k_{\rm f}}

\def\urms{u_{\rm rms}}
\def\urmsp{u^\prime_{\rm rms}}

\def\etat{\eta_{\rm t}}
\def\etatz{\eta_{\rm t0}}

\def\Beq{B_{\rm eq}}
\def\Btt{\overline{B}{}_\phi^{\rm rms}}

\def\half{{\textstyle{1\over2}}}

\newcommand{\chiSm}{\chi_{\rm m}^{\rm SGS}}

\begin{document}

\titlerunning{Turbulent transport coefficients of solar-like stars}
\authorrunning{Warnecke et al.}

\title{Turbulent transport coefficients in spherical wedge\\ dynamo
  simulations of solar-like stars}
\author{J. Warnecke\inst{1,2} \and M. Rheinhardt\inst{2} \and S. Tuomisto\inst{3}\and P. J.\ K\"apyl\"a\inst{4,2,1} \and M. J.\  K\"apyl\"a\inst{1,2}
   \and A. Brandenburg\inst{5,6,7,8}}
\institute{Max-Planck-Institut für Sonnensystemforschung,
  Justus-von-Liebig-Weg 3, D-37077 G\"ottingen, Germany\\
\email{warnecke@mps.mpg.de}\label{inst1} 
\and ReSoLVE Centre of Excellence, Department of Computer Science, Aalto University, PO Box 15400, FI-00\ 076 Aalto,
Finland \label{inst2}
\and Physics Department, Gustaf H\"allstr\"omin katu 2a, P.O. Box 64,
FI-00014 University of Helsinki, Finland\label{inst3}
\and Leibniz-Institut f\"ur Astrophysik Potsdam, An der Sternwarte 16,
D-14482 Potsdam, Germany\label{inst4}
\and Nordita, KTH Royal Institute of Technology and Stockholm University,
Roslagstullsbacken 23, SE-10691 Stockholm, Sweden\label{inst5}
\and Department of Astronomy, AlbaNova University Center,
Stockholm University, SE-10691 Stockholm, Sweden\label{inst6}
\and JILA and Department of Astrophysical and Planetary Sciences,
Box 440, University of Colorado, Boulder, CO 80303, USA\label{inst7}
\and Laboratory for Atmospheric and Space Physics,
3665 Discovery Drive, Boulder, CO 80303, USA\label{inst8}
}

\date{Received 15 January 2016 / Accepted 4 July 2017}
\abstract{}{%
We investigate dynamo action in global compressible solar-like
convective dynamos in the framework of mean-field theory.
}{%
We simulate a solar-type star in a wedge-shaped spherical shell, where the
interplay between convection and rotation self-consistently drives 
a large-scale dynamo.
To analyze the dynamo mechanism we apply the test-field method
for azimuthally ($\phi$) averaged fields to
determine the 27 turbulent transport coefficients of the electromotive force,
of which six are related to the $\alpha$ tensor.
This method has previously been used either in simulations in Cartesian
coordinates or in the geodynamo context and is applied here for the
first time to fully compressible simulations of solar-like dynamos.
}{%
We find that the $\phi\phi$-component of the $\alpha$ tensor does not follow
the profile expected from that of kinetic helicity.
The turbulent pumping velocities significantly alter the effective
mean flows acting on the magnetic field and therefore challenge the
flux transport dynamo concept.
All coefficients are significantly affected 
by dynamically important magnetic
fields. Quenching as well as enhancement are being observed.
This leads to a modulation of the coefficients with the activity
cycle.
The temporal variations are found to be comparable to the
time-averaged values and seem to be responsible for a nonlinear
feedback on the magnetic field generation.
Furthermore, we quantify the validity of the Parker-Yoshimura rule for
the equatorward propagation of the mean magnetic field in the present case.
}{}

\maketitle

\section{Introduction}

The magnetic field in the Sun undergoes a cyclic modulation with a
reversal typically every eleven years.
One of the prominent features indicating the corresponding activity variation are
sunspots visible on the solar surface.
In the beginning of the cycle they occur predominantly at higher latitudes, 
but appear progressively at lower latitudes as the cycle unfolds.
This is likely to be caused by a dynamo mechanism operating in the convection zone
below the surface, where, due to the interaction of 
highly turbulent flows and rotation, a large-scale magnetic field is generated which
propagates from high latitudes to the equator over the course of the cycle.
Increasing computing power and access to highly parallelized numerical codes has made
it possible to reproduce some of the features of the equatorward
propagating solar magnetic field in global three-dimensional
convective dynamo simulations
\citep[e.g.,][]{GCS10,KMB12,WKKB14,ABMT15,DWBF15,GSGKM15,KKOBWKP15}.
However, none of these are operating in the parameter regime of the Sun.
The ratios of advective (or inductive) and diffusive terms in the evolution
equations of fluid velocity, magnetic field, and specific entropy in
the Sun are several orders of magnitude larger than even in the highest
resolution simulations available today.
Nevertheless, these simulations are able to provide fundamental
insights into the dynamo mechanisms acting in them and therefore possibly also
in the Sun and solar-like stars.

As the flows in the solar convection zone are highly turbulent, a
large number of turbulence effects are able to operate.
For example, differential rotation is believed to be generated by
turbulent redistribution of angular momentum and heat
\citep[see e.g.,][]{R89}.
Mean-field theory has been successful in describing 
large-scale dynamos operating
in the Sun and other astrophysical objects \citep[e.g.,][]{BS05}.
Here, small-scale contributions to the magnetic field evolution are
parameterized in terms of the mean magnetic field via turbulent
transport coefficients \citep[e.g.,][]{KR80}, giving rise to, for example, the
well-known $\alpha$ effect \citep{SKR66}.
Using this approach made it possible to reproduce and understand some of
the key magnetic features observed in the Sun
\citep[e.g.,][]{CSD95,DC99,KKT06}.
However, we can only obtain an order of magnitude estimate for the
turbulent diffusion coefficient from solar observations. This means
that the turbulent transport coefficients are therefore
drastically simplified and/or
adjusted so that the resulting mean-field solutions reproduce the
observed properties of the large-scale magnetic field.

Combining the global convective dynamo simulations with the 
descriptive and potentially predictive power of the mean-field
approach is a promising path towards identifying and understanding astrophysical
dynamo mechanisms.
The first steps in determining the $\alpha$
coefficient and the
turbulent pumping velocity from local convection simulations with an
imposed-field method were made some time ago
\citep{BTNPS90,OSB01,OSBR02,KKOS06}.
The coefficients computed in the latter work
have also been used in global mean-field models \citep{KKT06}.
The main caveat of the imposed-field method is that it only allows the
$\alpha$ tensor and the pumping velocity to be obtained,
but no higher-order terms such as turbulent diffusion.
Furthermore, the mean magnetic field needs to be uniform to allow a
unique determination of these quantities which is violated by the
fact that the interaction of the imposed field with the flow
leads to additional mean constituents.
This can be avoided by resetting the magnetic field before significant
gradients develop \citep{OSBR02,HDSKB09}. Otherwise the values of the $\alpha$
coefficient can be very misleading \citep[][and references
therein]{KKB10}.

As an alternative, yielding also the turbulent diffusivity tensor, \cite{BS02}
and \cite{KOH06} used a multidimensional regression method 
which exploits the time-varying property of the mean fields.
A simplified version of it, which does not yield the
turbulent diffusivity tensor, and employs the
singular value decomposition, was used first
by \cite{RCGBS11} and later also by, for example, \cite{SCB13}
and \cite{ABT13}. 
\cite{SCD16} have recently relaxed this simplification.

\cite{SRSRC05,SRSRC07} developed a general and accurate
method to determine
the full tensorial representation of the turbulent transport coefficients 
for arbitrary velocity fields, in particular those
from global convective dynamo simulations, using so-called test fields.
This test-field method proved very successful in determining
the dynamo mechanisms in simulations of planetary interiors
\citep{SRSRC05,SRSRC07,SPD12,Schr11}, Cartesian convection
\citep[e.g.,][]{KKB09},
accretion discs \citep[e.g.,][]{B05AN,GP2015},
Roberts flows \citep[e.g.,][]{TA08,DBM13}, and other
setups \citep[see][and references therein]{SBS08,BRRS08,BCDHKR10}. 
In \cite{SPD11}, the method was applied to analyze the induction
mechanisms in stellar-type oscillatory dynamos found in a Boussinesq
model. Remarkably, it was shown that the $\Omega$ effect does not
fully explain the existence of a dynamo wave as the $\alpha$ tensor
alone already gives rise to one.
However, the $\alpha$ tensor alone produces an equatorward migration of the mean magnetic field,
whereas the addition of an $\Omega$ effect leads to poleward
migration.

In this work we apply the test-field method to fully compressible solar-like global
convective dynamo simulations determining full
tensorial expressions of the transport coefficients.

\section{Model and setup}
\label{sec:model}
For the simulations performed here,
the solar convection zone was modeled as a spherical
wedge defined in spherical polar coordinates by $0.7\,R\leq r\leq R$, 
$15^\circ\leq\theta\leq165^\circ$ 
and $0^\circ\leq\phi\leq
90^\circ$, where $r$ is the radial coordinate (with $R$ being the
radius of the star), $\theta$ is the colatitude and $\phi$ is the
longitude.
We solve the compressible magnetohydrodynamic (MHD) equations for the 
density $\rho$, the velocity $\uu$, the magnetic field
$\BB=\nab\times\AAA$ in terms of the vector potential $\AAA$, and the
specific entropy $s$.
The pressure is defined via the ideal gas equation $p=(c_p-c_V)\rho T$,
where  $c_p $ and $c_V$ are the specific heats at constant pressure
and constant volume, respectively, and $T$ is the temperature. 
We employ stress-free conditions for the flow on the latitudinal and
radial boundaries.
For the magnetic field, perfect conductor conditions
are applied on the lower radial and both latitudinal boundaries,
but radial field conditions $B_\theta=B_\phi=0$ on the upper radial
boundary.
The thermal properties of the systems are constrained by prescribing
the energy flux on the lower radial boundary and setting vanishing
energy fluxes on the latitudinal boundaries.
At the upper radial boundary, we apply a blackbody condition
for the temperature at $r=R$.
All quantities are assumed to be periodic in the $\phi$ direction.
The details of the models and their initial conditions can be found in
\cite{KMCWB13} and \cite{WKKB14,WKKB15} and will not be repeated here.

Our simulations are controlled by the fluid and magnetic Prandtl
numbers $\PrSGS=\nu/\chiSm$, $\Pr=\nu/\chi_{\rm m}$ , and $\Pm=\nu/\eta$,
respectively, where $\nu$ is the kinematic viscosity, $\chi_{\rm
  m}=K(r_{\rm m})/c_{\rm P} \rho_{\rm m}$ is the thermal diffusivity,
$\rho_{\rm m}$ is the density, and $\chiSm$ is the subgrid scale (SGS)
heat diffusivity, the latter three
evaluated at $r=r_{\rm m}\equiv0.85\,R$.
Furthermore, $\eta$ is the magnetic diffusivity, the Taylor number is given by
$\Ta=(2\Omega_0\Delta r^2/\nu)^2$, with $\Omega_0$ being the rotation rate of the
star, and the Rayleigh number by
$\Ra\!=\big(GM(\Delta r)^4/c_{\rm P} \nu\chiSm R^2 \big) (-{\rm
  d}s_{\rm hs}/{\rm d}r)_{r_{\rm m}}$
evaluated for
the thermally equilibrated hydrostatic state (hs) where $G$ is
Newton's gravitational constant, $M$ is the mass of the star, and
$\Delta r=0.3\,R$ is the depth of the convection zone.
Important diagnostic parameters of our simulations are the Coriolis
number, $\Co=2\Omega_0/\urms\kf$, along with the fluid and magnetic
Reynolds numbers, $\Rey=\urms/\nu\kf$ and $\Rm=\urms/\eta\kf$, respectively.
Here, $\kf=2\pi/\Delta r\approx21/R$ is the wavenumber of the largest
vertical scale in the convection zone and
$\urms=\sqrt{(3/2)\brac{u_r^2+u_{\theta}^2}_{r\theta\phi t}}$ is
the rms velocity without the $\phi$ component, which is dominated by
the differential rotation.
The 3/2 factor is employed so as to have a
diagnostic parameter comparable to that of earlier work \citep[e.g.,][]{SBS08}.

Our model is similar to Run~I of \cite{WKKB14}, except that there 
$\PrSGS=2.5$ instead of $2$.
The run is therefore almost identical to Runs~B4m and C1 of \cite{KMB12} and
\cite{KMCWB13}, respectively.
A very similar run was also discussed in \cite{WKKB15} (their Run~A1)
and \cite{KKOWB17} (their Run~D3),
with slightly different stratification.
We also refer to 
the  hydrodynamic counterpart
of our model
labeled as HD (instead of MHD).
An overview of the parameters of the runs is shown in \Tab{runs}.
A comparison of typical parameters of solar-like dynamo simulations
by other authors can be found in Appendix~A of \cite{KKOWB17}.
For the saturated state of the run,
the radial profiles of $\theta\phi$--averaged
 temperature $T$, density $\rho$, and turbulent rms velocity can be found in Fig. 1 of
\cite{WKKB14} (with Run~I).
On average, the density contrast between bottom and top is roughly 22.
\begin{table}[t!]\caption{
Summary of Runs.
}\vspace{12pt}\centerline{\begin{tabular}{lccccccc}
Run & $\PrSGS$ &$\Pr$&$\Pm$&$\Ta$[$10^{8}$]& $\Ra$[$10^{7}$] &$\Co$ &$\Rey$ \\[.8mm]   
\hline
\hline\\[-2.5mm]
MHD &  2.0 & 60 & 1    & 1.3 &  4.2 & 8.3 & 34 \\   
HD    &  2.0 & 60  & --  & 1.3 &  4.2 & 8.2 & 34    \\[.5mm]
\hline
\hline
\label{runs}\end{tabular}}\tablefoot{
Second to sixth columns: input parameters.
Last two columns: diagnostics computed from the saturated states of
the simulations.
The Rayleigh number is around $100$ times the critical value for convection.
}
\end{table}
The slightly lower $\PrSGS$ compared to Run~I of \cite{WKKB14} does not
have a strong influence on differential rotation and magnetic field
evolution \citep{WKKB15}.
Even a simulation with $\PrSGS=1$ as in \cite{KKOBWKP15} shows no
significant qualitative difference.
The adopted Rayleigh number of $4.2\times10^7$ is around $100$ times the
critical value.

Throughout this paper, we will invoke the mean-field approach, 
within which we decompose quantities such as $\BB$ and $\uu$ into mean
and fluctuating parts, $\meanBB$ and $\flucbb=\BB-\meanBB$ as well as
$\meanUU$ and $\flucuu=\uu-\meanUU$, respectively.
We define the mean as the azimuthal (i.e., $\phi$)
average.
Thus, as is well known, dynamos with azimuthal order $m \ge 1$, as 
found in \cite{Coletal14}, cannot be described by such averaging.
Here we often use additional temporal or spatial
averages denoted as $\brac{.}_{\rm \xi}$, with ${\rm\xi}=t,r,\theta$.
One important quantity defined this way is the meridional distribution of the
turbulent velocity 
$\urmsp(r,\theta)={\left\langle\,\overline{{\bm
        u}^{\prime\,2}}\,\right\rangle_t}{}^{\!\!1/2}$
which takes all velocity components into account.
When presenting the results, we often use a normalization for the
transport coefficients motivated by the
first-order-smoothing approximation (FOSA), employing
$\alpha_0=\urmsp/3$ and $\etatz=\tau u^{\prime\,2}_{\rm rms}/3$ with
an estimate of the convective turnover time
$\tau=H_p\alpha_{\rm MLT}/\urmsp$, where 
$H_p=-(\partial \ln\mean{p}/\partial r)^{-1}$ is the pressure scale
height and $\alpha_{\rm MLT}$ is the mixing length parameter chosen here to
have the value $5/3$. 
We note that these normalization quantities depend on radius and
latitude.

The results below are either presented as normalized quantities or in
physical units by choosing a normalized rotation rate $\Ot$
=$\Omega_0/\Omega_{\odot}$=5, where
$\Omega_{\odot}=2.7\times10^{-6}\,$s$^{-1}$ is the solar rotation rate,
and assuming the density at the base of the convection zone
($r=0.7R_\odot$) to have the solar value $\rho=200$ kg~m$^{-3}$; 
see more details and discussion about the relation of the
simulations to real stars in \cite{KMCWB13,KMB14}, \cite{WKKB14} and
\cite{KKOBWKP15}.
The simulations were performed with the {\sc Pencil
  Code}\footnote{\tt{http://github.com/pencil-code/}}, which uses a
high-order finite difference method for solving the compressible
equations of MHD.

\section{Test-field method}
\label{sec:testfield}

\subsection{Theoretical background}
\label{sec:theory}
We consider the induction equation in the mean-field approach
\begin{equation}
{\partial \meanBB\over\partial
  t}=\nab\times(\meanUU\times\meanBB+
\overline{\flucuu\times\flucbb}) - \nab\times\eta\nab\times\meanBB,
\label{eq:mind}
\end{equation}
where 
\begin{equation}
\overline{\flucuu\times\flucbb}=\EMF
\label{eq:emf}
\end{equation}
is the mean (or turbulent)
electromotive force arising from the correlation of
the fluctuating velocity and magnetic fields.
We note that \eq{eq:mind} is an exact equation in MHD, where no
assumptions have been made except that the average must obey the
Reynolds rules, which the azimuthal average does.
At this stage no scale separation is required.
The $\EMF$ can be expanded in terms of the mean magnetic field
$\meanBB$,
\begin{equation}
\EMF=\aaaa\cdot\meanBB+\bb\cdot\nab\meanBB+\ldots
\label{eq:emf1}
,\end{equation}
where in the following we truncate the expansion after the first-order
spatial derivatives of $\meanBB$ and disregard any temporal
derivatives. 
This, however, does require scale separation, hence only the effects
of the magnetic field at the larger scales will be captured by this
approach.
Likewise, a proper representation of $\EMF$ by \eq{eq:emf1} can be
expected only for slowly varying mean magnetic fields.
We emphasize that this is not a principal restriction and that it has
been relaxed in earlier applications of the test-field method
\citep{BRS08,HB09,CMRB11, RB12}.
In \eq{eq:emf1}, $\aaaa$ and $\bb$ are tensors of rank two and three,
respectively.
Dividing these, as well as the derivative tensor $\nab\meanBB$ into
symmetric and antisymmetric parts, we can rewrite \eq{eq:emf1} as
(neglecting higher order terms indicated by \ldots)
\begin{equation}
\EMF=\aalpha\cdot\meanBB+\ggamma\times\meanBB 
-\bbeta\cdot(\nab\times\meanBB) 
-\ddelta\times(\nab\times\meanBB) 
-\kkappa \cdot(\nab\meanBB)^{(s)},
\label{eq:emf2}
\end{equation}
where $\aalpha$ is the symmetric part of $\aaaa$ giving rise to the
$\alpha$ effect \citep{SKR66}, $\gamma_i=-\epsilon_{ijk}
a_{jk}/2$ characterizes the antisymmetric part of $\aaaa$ and
describes changes of the mean magnetic field due to an effective
velocity $\ggamma$ (also:  ``turbulent pumping'')
\citep[e.g.,][]{OSBR02}, $\bbeta$ is the symmetric part of the rank two
tensor acting upon $\nab\times\meanBB$, which characterizes the
turbulent diffusion, $\ddelta$ quantifies its antisymmetric
part and enables what is known as the R\"adler effect \citep{KHR69}, 
$(\nab\meanBB)^{(s)}$ is the symmetric part of the derivative tensor
and $\kkappa$ is a rank-three tensor, whose interpretation
has yet to be established.
Detailed descriptions of these tensors are provided in
Sections~\ref{sec:alph}, \ref{sec:pump} and \ref{sec:diff}.

Calculating these transport coefficients will enable the
identification of the physical processes which are responsible for the
evolution and generation of the mean magnetic field.
The test-field method \citep{SRSRC05,SRSRC07,SPD12} is one way
to calculate these coefficients from global dynamo simulations.  
To compute $\EMF$, we solve
\begin{equation}
\begin{aligned}
{\partial\fluc{\bb}_{\rm T}\over\partial t}=&\,\nab\times\left(\flucuu\times\meanBB_{\rm T}  +
\meanUU\times\fluc{\bb}_{\rm T} + \flucuu\times\fluc{\bb}_{\rm T} -
\overline{\flucuu\times\fluc{\bb}_{\rm T}}\,\right) \\
&- \nab\times\eta\nab\times\fluc{\bb}_{\rm T}
\end{aligned}
\label{eq:testpr}
\end{equation}
for  $\fluc{\bb}_{\rm T}$
with a chosen test field $\meanBB_{\rm T}$,
while taking $\meanUU$ and $\flucuu$ from the global simulation (the ``main run"),
and employ \eq{eq:emf}.
By choosing nine linearly independent test fields, we
have a sufficient number of realizations of \eq{eq:emf1}
to solve for all coefficients of \eq{eq:emf2}.
A detailed description and discussion, in particular for spherical
coordinates, can be found in \cite{SRSRC05,SRSRC07}.

The test-field method in the presented form is only valid in the 
absence of a ``primary magnetic turbulence'',  that is, if the
magnetic fluctuations $\flucbb$ vanish for
$\meanBB\equiv\boldsymbol{0}$.
However, for sufficiently high magnetic Reynolds numbers, a small-scale
dynamo may exist which creates magnetic fluctuations also in the
absence of $\meanBB$.
For the simulation considered here, this can be ruled out:
a test run, where $\meanBB$ has been removed in each time step shows no
magnetic field growth.

\subsection{Implementation}
\label{sec:implem/}
The implementation of the test-field method follows the lines
described in \cite{SRSRC05,SRSRC07}: The nine test fields were specified such
that each has only one non-vanishing spherical component and is either
constant or depends linearly on $r$ or $\theta$, see Table~1 of
\cite{SRSRC07}.
We note that some of these fields are not solenoidal or become irregular
at the axis, and that none of them obey the boundary conditions posed 
in the main run, but these properties have been shown by the same
authors not to exclude the suitability of such fields.
Clearly, they form a linearly independent function system.
The nine test problems resulting from \eq{eq:testpr} are integrated along with the
main run while simultaneously forming the mean electromotive forces $\EMF$
out of their solutions and inverting nine disjoint equation systems of rank
three to obtain three of the 27 transport coefficients by each. 
At higher $\Rm$, some of the eigensolutions of the homogeneous parts
of the test problems can become unstable.
To suppress their influence, the test solutions are  re-initialized to zero in regular time intervals 
\citep{MKTB09,HDSKB09}.
Their length is typically chosen to be at least 30 turnover
times.  As the transport coefficients are also required to reflect the
temporal changes in the turbulence due to the magnetic cycle, an upper
bound is set by a sufficiently small fraction (say, one tenth) of the
cycle period.

To obtain the coefficients
$\tilde{a}_{\mu\lambda}$, $\tilde{b}_{\mu\lambda r} $ and
$\tilde{b}_{\mu\lambda \theta}$ in the non-covariant relation,
\begin{equation}
\emf_\mu   =  \tilde{a}_{\mu\lambda} \meanB_\lambda +
\tilde{b}_{\mu\lambda r} \frac{\partial\meanB_\lambda}{\partial r} +
\tilde{b}_{\mu\lambda \theta} {1\over r}\frac{\partial \meanB_\lambda}{\partial
  \theta}, \quad   \lambda=r,\theta,\phi, 
\label{eq:EMFmu}
\end{equation}
we filter out the initial, transient
epochs and those contaminated by
the unstable eigensolutions,
and perform a reliability check of
statistical (quasi-) stationarity. The (covariant) coefficient tensors in \eq{eq:emf2} are then obtained
from the non-covariant ones employing the relations (18) of
\cite{SRSRC07}.
We note that their sign conventions for $\aalpha$ and $\ggamma$
are different from ours.
The implementation has been validated using a simple model of a
forced turbulent dynamo and comparing it with 
a corresponding mean-field model;
see \App{sec:appB}.
We have also verified it using a 
stationary laminar flow;
see \App{sec:appC}.

\section{Results}

In \Secss{sec:alph}{sec:diff} we focus on the analysis of the
time-averaged transport coefficients,
for simplicity and compactness leaving out $\bra{\cdot}_t$ indicating
time averaging, while in
\Sec{sec:var} we discuss the variations in time.
In \Secs{sec:mag_quen}{sec:cycl} we investigate the magnetic quenching
and cyclic variation of the transport coefficients due to the mean
magnetic field.
In \Sec{sec:park} we discuss the mean magnetic field
propagation by applying a similar technique as in \cite{WKKB14}.
Finally, in \Sec{sec:bs} we compare the results from the
test-field method with results obtained from the multidimensional
regression method used by \cite{BS02} and later by, for example, \cite{RCGBS11}, \cite{ABMT15}, and \cite{SCD16}.

\subsection{Meridional profiles of $\aalpha$}
\label{sec:alph}
\begin{figure}[t!]
\begin{center}
\includegraphics[width=1.00\columnwidth]{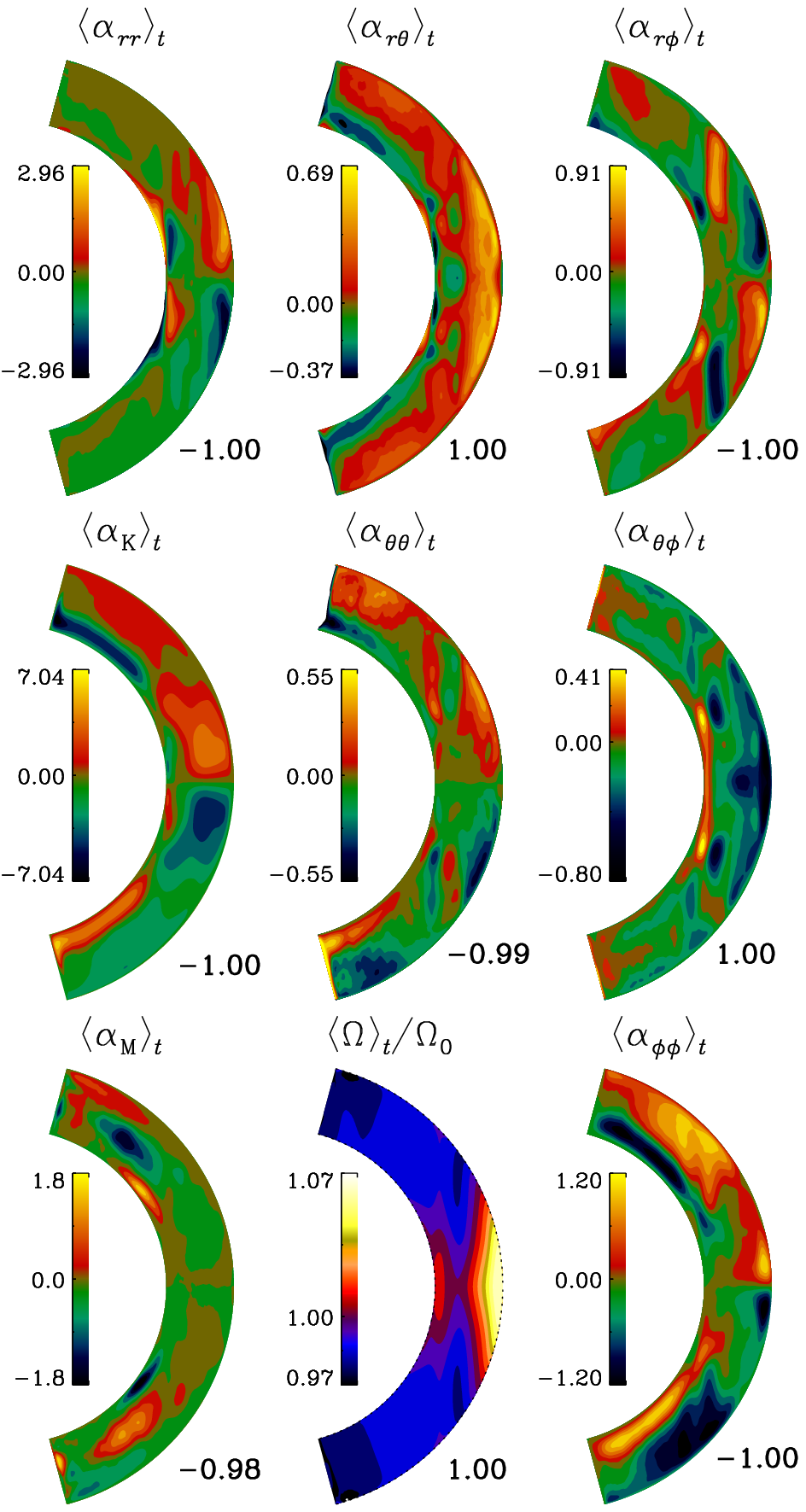}
\end{center}\caption[]{
 Components of $\aalpha$ and $\alpha_{\rm K,M}$, normalized by
 $\alpha_0=\urmsp/3$, and normalized differential rotation
 $\Omega/\Omega_0$;
all quantities are time averaged. 
Numerals  at the {\it bottom right} of {\it each panel}: 
overall parity $\tilde{P}$, see \eq{eq:parity}.
}\label{alpij}
\end{figure}
\begin{figure}[t!]
\begin{center}
\includegraphics[width=1.\columnwidth]{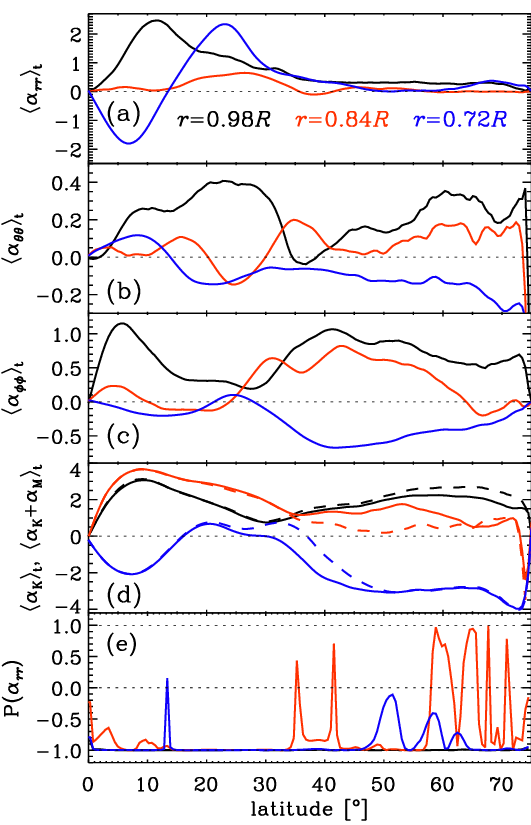}
\end{center}\caption[]{
Time-averaged main-diagonal components of $\aalpha$ (a--c) together
with $\alphaK$, $\alphaK$+$\alphaM$ (d) and the parity $P(\alpharr)$ 
(see \eq{eq:parity}) (e) over latitude 90$^\circ-\theta$ in
the northern hemisphere and for three different radii:
$r=0.98\,R$ (black), $r=0.84\,R$ (red), $r=0.72\,R$ (blue).
Solid and dashed lines in d:
 $\alphaK$ and  $\alphaK+ \alphaM$, respectively.
 Values in a-d are normalized by $\alpha_0=\urmsp/3$.
}\label{alplat}
\end{figure}
In \Fig{alpij} we plot the time averages of all components of
$\aalpha$.
All three diagonal components of 
$\aalpha$ are mainly positive in the
north and negative in the south, but have a sign reversal in the lower
layers of the convection zone (except $\alpharr$).
This behavior is similar to that of $\alpha$ for isotropic and
homogeneous turbulence in the low-dissipation limit \citep{PFL76} via
\begin{equation}
\alpha=-{\tau\over
  3}\left(\overline{\fluc{\oo}\cdot\flucuu}
  -\overline{\fluc{\jj}\cdot\flucbb}/\meanrho \right)
\equiv\alphaK+\alphaM,
\label{eq:aKaM}
\end{equation}
where $\alphaK$ and $\alphaM$ are the kinetic and magnetic $\alpha$
coefficients, respectively, $\fluc{\oo}=\nab\times\flucuu$ is the
fluctuating vorticity, $\overline{\fluc{\oo}\cdot\flucuu}$ is the small-scale
kinetic helicity, $\fluc{\jj}=\nab\times\flucbb/\mu_0$ is the
fluctuating current density,
$\overline{\fluc{\jj}\cdot\flucbb}$ is the small-scale current
helicity, and $\meanrho$ is the mean density.
For a direct comparison we plot the meridional distributions of
$\alphaK$ and $\alphaM$ in \Fig{alpij} as well as the
latitudinal profiles  of the diagonal components of $\aalpha$ together
with those of $\alphaK$ and $\alphaK$+$\alphaM$ at three different
depths in \Fig{alplat}.

It turns out that $\alpharr$ is the strongest of all components of $\aalpha$,
in particular in concentrations near the surface at low latitudes; see
\Figs{alpij}{alplat}.
The same has been found previously for Cartesian shear flows using
both multidimensional regression methods \citep{BS02,KOH06} as well as
the test-field method \citep{B05AN}.
Unfortunately, a comparison with \cite{KKB09}, where transport
coefficients for convection in a Cartesian box have been obtained by
the test-field method, is not possible as $\alpharr$ was not determined
there.
In the middle of the convection zone, $\alpharr$ is much weaker than
above and below; but compared to the other components of $\aalpha$
 the values are still high or similar ($\alphapp$).
The latitudinal dependency shows a steep decrease from low to high
latitudes.

Next, $\alphatt$ is around six and two times weaker than
$\alpharr$ and $\alphapp$, respectively, and shows multiple sign
reversals on cylindrical contours; see \Fig{alpij}.
A region of negative (positive) $\alphatt$ 
at mid-latitudes in the northern (southern)
hemisphere coincides with a local minimum of the rotation rate 
$\Omega(r,\theta)=\Omega_0+\brac{\mean{U_\phi}}_t/r\sin\theta$ as seen in
\Fig{alpij} and a maximum of negative radial and latitudinal shear
($\partial_r \Omega <0$, $\partial_\theta\Omega<0$); see 
bottom row of \Fig{effects}.
In the results of \cite{SRSRC07,SPD12} and \cite{Schr11}, $\alpharr$ and
$\alphatt$ have opposite signs in comparison to the present work.
In their results, the $rr$ and $\theta\theta$ components are negative (positive)
near the surface in the northern (southern) hemisphere and positive
(negative) deeper down and therefore do not show a pattern similar to
$\alpha_{\phi\phi}$ and $\alphaK$ as in our work.
Furthermore, their $\alpharr$ and $\alphatt$ 
are around four times weaker
than $\alphapp$, whereas in our work $\alpharr$ is the largest.

Further, $\alphapp$ shows concentrations at low and mid to high
latitudes near the surface, but also in deeper layers, where its sign
is opposite to that near the surface.
This sign reversal with depth is most pronounced in $\alphapp$, but
also visible in $\alphatt$.
The meridional profile of $\alphapp$ is roughly similar to that of
$\alphaK$, although its strength is smaller, see \Figs{alpij}{alplat}.
This is consistent with the finding of \cite{SRSRC07}, where
$\alphapp$ was similar to $\alphaK$.
Therefore a mean-field model using $\alphaK$ was not able to reproduce
the magnetic field, while a model using $\alphapp$ was.
The latitudinal dependencies of $\alphapp$ and $\alphaK$ follow neither
a typical cosine distribution as found by, for example, \cite{KKOS06} for
moderate rotation, nor a $\sin\theta\cos\theta$ distribution as often
assumed in Babcock-Leighton dynamo models \citep[e.g.,][]{DC99}.
In \cite{KKB09}, an increase of $\alpha_{\theta\theta}$ and $\alpha_{\phi\phi}$
from the equator to the poles was found, but the functional form is
not clear.

The off-diagonal components of $\aalpha$ have strengths similar to
$\alphatt$ and $\alphapp$ and are therefore significantly weaker than
$\alpharr$.
Though of opposite sign, $\alpha_{r\theta}$ and $\alpha_{\theta\phi}$ have similar
equatorially symmetric profiles,
with positive and negative values, respectively, in the
upper $\gtrsim$ 75\% of the convection zone
-- in particular below mid-latitudes.
Finally, $\alpha_{r\phi}$ is similar to $\alphatt$, but the sign
reversal in the region of minimum $\Omega$ at mid-latitudes
is more pronounced 
in $\alpha_{r\phi}$
and at high latitudes the sign is the same.
 While $\alpha_{r\theta}$ and $\alpha_{\theta\phi}$ have the same sign,
$\alpha_{r\phi}$ has the opposite sign compared to \cite{SRSRC07}.
In general, the coefficients obtained in the present paper are less
``cylindrical" than in the work of \cite{SRSRC07,SPD12} and
\cite{Schr11}, probably because in our simulation, rotation is slower and
convection is more supercritical.

Inspection by eye already suggests that the components of $\aalpha$
are almost fully equatorially symmetric or
antisymmetric\footnote{For the special solutions of the full MHD
problem in a model with only $r$-dependent coefficients, 
  given by equatorially symmetric velocity, density and
  entropy with an either symmetric or antisymmetric magnetic field,
  it can be shown that the main diagonal components of $\aalpha$,  as
  well as $\alpha_{r\phi}$  are antisymmetric, all
  other components symmetric.}.
In order to study these symmetries quantitatively we define the
pointwise parity of a quantity, for example, $\alpha_{ij}$, as
\begin{equation}
P(\alpha_{ij})={
\left(\alpha_{ij}^{\rm s}\right)^2-\left(\alpha_{ij}^{\rm a}\right)^2\over   
\left(\alpha_{ij}^{\rm s}\right)^2+\left(\alpha_{ij}^{\rm a}\right)^2},
\label{eq:parity}
\end{equation}
where $\alpha_{ij}^{\rm s,a}(r,\theta)=\half\left[\brac{\alpha_{ij}(r,\theta)}_t
  \pm\brac{\alpha_{ij}(r,\pi-\theta)}_t\right]$ are the equatorially
symmetric and antisymmetric parts of $\alpha_{ij}$, respectively.
In the bottom panel of \Fig{alplat}, we exemplarily plot $P(\alpharr)$.
As expected, its value is in most of the meridional plane $-1$,
corresponding to antisymmetry. 
This is particularly valid near the surface (black line).
The locations, where the parity is different coincide with
values of $\alpharr$ being close to zero, and are of low significance.
All other $\aalpha$ components show the same small deviations from the 
pure parity state $P(\alpha_{r\theta})=P(\alpha_{\theta\phi})=1$ and
$P(\alphatt)=P(\alphapp)=P(\alpha_{r\phi})=-1$.
To describe the overall parity of a coefficient by a single number
$\tilde{P}$, we
also employed \eq{eq:parity} with additional volume integrations in
numerator and denominator; see
Figs.~\ref{alpij}, \ref{turb}, \ref{betij}, \ref{kapijk}, and \ref{bs} for these
values.
For $\aalpha$ we have $|\tilde{P}|\gtrsim 0.99$ which is consistent
with the almost pure overall equatorial symmetry of the velocity
field\footnote{We note that the symmetric
(antisymmetric) part of a vector field $\VV$ is constituted by the
  symmetric (antisymmetric) parts of $V_{r,\phi}$, but the
  antisymmetric (symmetric) part of $V_\theta$.}: $\tilde{P}(\mean{U}_r)=0.99$,
$\tilde{P}(\mean{U}_\theta)=-0.99$, $\tilde{P}(\mean{U}_\phi)=1.00$
and $\tilde{P}(\urmsp)=1.00$.

\subsection{Magnetic field generators}
\label{sec:mag_gen}

\begin{figure*}[t!]
\vspace{10.4cm}
\sidecaption
\includegraphics[width=.84\columnwidth]{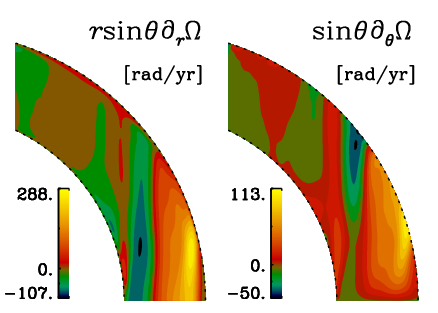}
\caption[]{
The dominant effects in the mean magnetic field evolution.\\
{\it Top row, from left to right:} $\Omega$ effect
$r\sin\theta\,\meanBBp\cdot\nab\Omega$, toroidal $\alpha$ effect
$(\nab\times\aalpha\cdot\meanBB)_{\phi}$, toroidal turbulent diffusion
$(\nab\times\bbeta\cdot\nab\times\meanBB)_{\phi}$, 
mean radial and latitudinal field $\mean{B}_r$, $\mean{B}_\theta$.
White: Field lines of mean poloidal field $\meanBBp$.  
{\it Middle row:} Radial and latitudinal $\alpha$ effect
$(\nab\times\aalpha\cdot\meanBB)_{r,\theta}$, radial and latitudinal
turbulent diffusion $(\nab\times\bbeta\cdot\nab\times\meanBB)_{r,\theta}$,
mean toroidal field $\mean{B}_\phi$.
{\it Bottom row:} Components of $r\sin\theta\nab\Omega$.
We note that the effects are computed with $\meanBB$ at the halfway point of 
a typical activity cycle with positive $\meanBBt$ (cf. \Fig{avcyl}, top),
but with the time-averaged transport coefficients (see
\Sec{sec:mag_gen}).
}\label{effects}
\end{figure*}

To investigate the relative importance of the main contributions to
mean magnetic field evolution in detail, we plot those from
the $\Omega$ and
$\alpha$ effects as well as 
from
the turbulent diffusion in
\Fig{effects} along with the components of $\meanBB$ and the shear.
Contributions from the meridional circulation have turned out to be
significantly weaker; see the dynamo number calculations in
\cite{KMCWB13} and \cite{KKOBWKP15}.
We also do not show the contributions related to
$\ggamma$, $\ddelta$ or $\kkappa$.
Here, $\aalpha$, $\bbeta$ and shear have been
time-averaged over all cycles in the saturated stage.
For $\meanBB,$ we first constructed a typical magnetic cycle by
folding all magnetic cycles on top of one another and averaging.
Then we selected the instant at the half of the activity cycle with positive
toroidal magnetic field near the surface
at low latitudes and used the corresponding $\meanBB$ for the calculations.
 
We employ here the poloidal-toroidal decomposition of the mean magnetic
field with $\meanBBp=\mean{B}_r\eee_r+\mean{B}_\theta\eee_\theta$,
$\meanBBt=\mean{B}_\phi\eee_\phi$ and $\eee_i$ being the unit vector
in the direction $i$.
The $\Omega$ effect shears the mean poloidal field, generating mean
toroidal field via $\meanBBp\cdot\nab\Omega$ (top row of \Fig{effects}).
At mid latitudes we find two distinct contributions: Outside the
tangent cylinder\footnote{The cylinder aligned with the rotation axis
  and tangent to the sphere bounding the domain from below.}
 the negative radial shear (see bottom row of \Fig{effects})
generates a negative toroidal field from the positive radial field.
Further away from the tangent cylinder the sign of the radial shear
changes and it produces positive toroidal field.
These two regions of field production coincide well with those of
strong $\meanBBt$ as shown in the middle row of \Fig{effects}.
Inside the tangent cylinder, the positive latitudinal shear generates positive toroidal field
again, but weaker than the radial shear does,
and we find a corresponding region of positive toroidal field.
Beside these pronounced regions, there is also negative toroidal field
production near the surface due to radial shear at high latitudes.
However, for these regions, there seems to be no clear relation to
toroidal field concentrations at this instant of the cycle.

At the same time, the $\alpha$ effect can also generate toroidal
magnetic field via $(\nab\times\aalpha\cdot\meanBB)_\phi$; see 
the top row of \Fig{effects}.
This involves radial derivatives of $\alpha_{\theta i}\mean{B}_i$  
and latitudinal derivatives of $\alpha_{r i}\mean{B}_i$.
One finds that the $\alpha$
effect generates toroidal field of the same sign as the $\Omega$
 effect at mid latitudes just outside the tangent cylinder, therefore
enhancing its negative toroidal field production.
However, the contribution from $\aalpha$ has only one third of the
strength of the $\Omega$ effect.
Directly next to this region, further away from the tangent cylinder,
the $\alpha$ effect generates  positive toroidal field similar to the
$\Omega$ effect, but again weaker.
Additionally, the $\alpha$ effect is strong close to the surface,
producing positive toroidal field at mid- to high latitudes
(at high latitudes mostly opposite to the $\Omega$ effect).
Close to the surface at high latitudes it is comparable to the $\Omega$
effect, while at low latitudes in the upper half of the convection
zone it is locally stronger.
This is suggestive of an $\alpha^2$ dynamo being dominant near the
surface, while the $\alpha^2\Omega$ dynamo dominates in deeper layers
which
is supported by phase relation measurements of \cite{KMCWB13}:
Equatorward migration is explained by an $\alpha^2\Omega$ dynamo
wave in the bulk of the convection zone \citep{WKKB14}, while near the
surface the phase relation is consistent with an $\alpha^2$ dynamo.
Here, \cite{KMB12} found a high-frequency dynamo mode
in addition to the main mode as discussed in
detail in \cite{WKKB14} and \cite{KKOBWKP15} for similar runs.
However, the hypothetical $\alpha^2$ dynamo has to be verified in a
mean-field model.
This result seems to be consistent with the conclusion presented in
\cite{SPD11}, where the authors show that $\alpharr$  also has a
strong contribution to the magnetic field evolution and actually sets
the cycle period.
However, in our simulation, the main generator of toroidal magnetic
field is still the $\Omega$ effect, in particular for the equatorward
migrating field.
The toroidal $\alpha$ effect also shows negative contributions at high
latitudes in the upper half of the convection zone, where it takes part
in the generation of toroidal field;
see top and middle rows
of \Fig{effects}.

In the third panel of the top row of \Fig{effects} we plot the toroidal contribution 
of the turbulent diffusion using the full $\bbeta$, which is
discussed in detail in \Sec{sec:diff}.
It has roughly the same strength as the main toroidal field generators,
but shows the opposite sign in the mid-latitude regions of strong field production.
However, it exhibits far stronger spatial variations than the
corresponding $\Omega$ and $\alpha$ effects and thus does not
match up well with the production terms.
The structures at small spatial scales can be taken as indications of poor scale separation
between mean and fluctuating quantities, pointing to the need of scale-dependent
transport coefficients; see \Sec{sec:theory}.
We note here that a simplified treatment with $(\eta+
\etatz)\big(\Delta \meanBB\big)_\phi$ yields excess values by a factor of three
and an incorrect distribution.

We also investigated the $\alpha$ effect generating $\meanBBp$ from
$\meanBBt$ via $\big(\nab\times\aalpha\cdot\meanBB\big)_{r,\theta}$;
see middle row of \Fig{effects}.
The $\alpha$ effect contribution to the poloidal field seems to be
weaker than that of the $\Omega$ effect to the toroidal one
giving an indication of the weaker poloidal than toroidal field at mid
and lower latitudes.
However,  in the deeper convection zone at mid to
high latitudes, $\mean{B}_\theta$ is comparable with $\mean{B}_\phi$.
The positive radial contribution of the $\alpha$ effect fits well with the 
positive radial field at mid latitudes outside the tangent
cylinder. However, it generates negative radial field slightly
toward higher latitudes, which has no correspondence in the radial field.
At high latitudes the radial $\alpha$ effect shows strong spatial variations
producing mostly positive radial field.
At low latitudes, one finds small patches of positive and negative
radial field generation.

The latitudinal $\alpha$ effect also shows field generation at
mid-latitudes outside the tangent cylinder coinciding well with
positive latitudinal field there.
Interestingly, the latitudinal $\alpha$ effect is particularly   
pronounced close to the surface, similar to the azimuthal $\alpha$
effect.
This might be an indication of a possible $\alpha^2$ dynamo
operating in these regions.
However, the radial $\alpha$ effect does not show such a profile.
For the poloidal field, we also calculated the turbulent diffusion;
see  middle row of \Fig{effects}.
As for the toroidal field, the diffusion shows strong spatial
variations.
At the surface the latitudinal diffusion seems to counteract the
latitudinal $\alpha$ effect.
The production sites of toroidal field seem to match with the
contributions shown in \Fig{effects}, but there are still some differences.
The poloidal field production, however, cannot be linked directly to the
shown production and diffusion terms, due to the phase-shift between
$\meanBBt$ and $\meanBBp$ as shown in \cite{WKKB14}.
That is why the latitudinal $\alpha$ effect produces negative latitudinal
field, where this is positive.
Furthermore, there are additional effects in the mean
electromotive force (\Secss{sec:pump}{sec:cycl}), which together
with the meridional flow affect the mean field.

\begin{figure}[t!]
\begin{center}
\includegraphics[width=1.00\columnwidth]{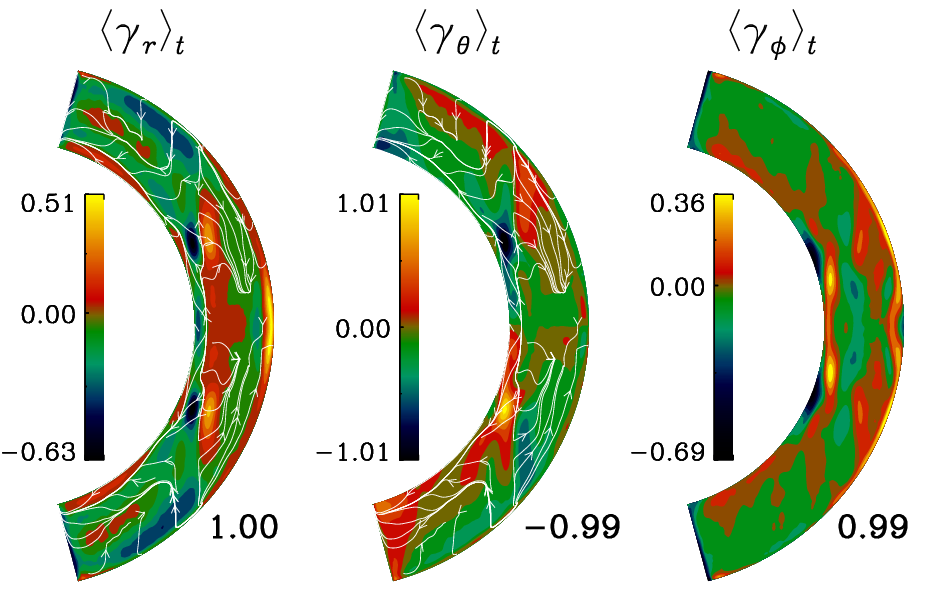}
\end{center}\caption[]{
Components of $\ggamma$ normalized by $\alpha_0=\urmsp/3$.
White streamlines: meridional pumping velocity
$\gamma_r
\eee_r+\gamma_\theta \eee_\theta$.
Numerals  at the bottom right of each panel: 
overall parity $\tilde{P}$, see \eq{eq:parity}.
}\label{turb}
\end{figure}
\begin{figure*}[t!]
\begin{center}
\includegraphics[width=.85\textwidth]{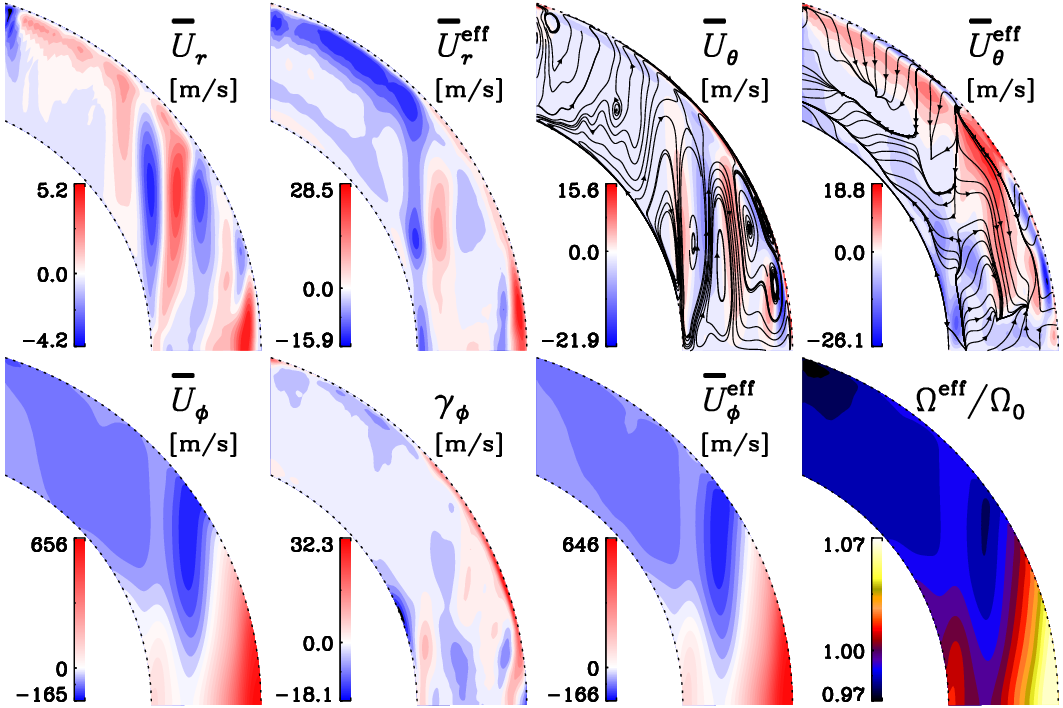}
\end{center}\caption[]{{\it Top, from left to right:}
Time-averaged radial and effective radial flow $\meanU_r$,
$\meanU{}^{\rm eff}_r = \meanU_r +\gamma_r$, latitudinal and effective
latitudinal flow $\meanU_\theta$, $\meanU{}^{\rm eff}_\theta
=\meanU_\theta+\gamma_\theta$. {\it Bottom:} azimuthal flow  $\meanU_\phi$,
$\gamma_\phi$, effective azimuthal flow $\meanU{}^{\rm
  eff}_\phi=\meanU_\phi+\gamma_\phi$ and effective differential
rotation $\Omega^{\rm eff}=\meanU{}^{\rm eff}_\phi/r \sin\theta +
\Omega_0$.
Solid lines in top row with arrows: flow lines of $\meanUUp$, $\meanUU{}^{\rm
  eff}_{\rm pol}$.
}\label{effvel}
\end{figure*}

\subsection{Turbulent pumping}
\label{sec:pump}

\begin{figure}[t!]
\begin{center}
\includegraphics[width=.98\columnwidth]{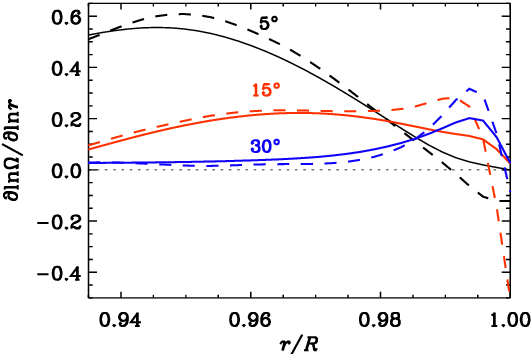}
\end{center}\caption[]{
Double logarithmic radial gradient of time-averaged rotation rate,
$\partial\ln\Omega/\partial\ln r$ (solid), and effective rotation rate,
$\partial\ln\Omega^{\rm eff}/\partial\ln r$ (dashed), at 5$^\circ$
(black), 15$^\circ$ (red) and 30$^\circ$ (blue) latitude.
}\label{NSsh}
\end{figure}

In \Fig{turb}, we plot the time-averaged components of the turbulent
pumping velocity $\ggamma$.
The latitudinal component
$\gamma_\theta$ is the strongest with extrema
of around $30\%$ of the turbulent velocity $\urmsp$.
The other components have a half ($\gamma_r$) and a third
($\gamma_\phi$) of the values.
Hence all components are comparable to the off-diagonal
components of $\aalpha$, while being three 
($\gamma_\theta$) to ten ($\gamma_\phi$) 
times weaker than
$\alpharr$.

For $\gamma_r$ there are two distinct regions:
Inside and close to the tangent cylinder, $\gamma_r$ is mainly negative
(except at high latitudes in the middle of the convection zone), whereas
further outside it is mainly positive.
Close to the surface, at low and mid-latitudes,
$\gamma_r$ is positive indicating outward
pumping of magnetic field.
This is at odds with downward pumping
found in local simulations of the near-surface layers
\citep[e.g.,][]{NBJRRST92,OSBR02},
but in agreement with what is expected from SOCA or even earlier
calculations, where the pumping is proportional to the negative gradient of  the turbulence
intensity ($\ggamma\approx-\nab\etat/2$) \citep[e.g.,][]{K91}.
In a study using scale-dependent
one-dimensional test fields in Cartesian convection simulations,
\cite{KKB09} found that downwards pumping occurs only for uniform
($k=0$) mean fields and outward pumping for all other field scales
($k>0$).
We find, however, that at high latitudes near the surface downward
radial pumping dominates.
The difference between high and low latitudes can be 
due to rotational influence.
The negative extrema of $\gamma_r$ occur near the bottom of the convection zone
close to the tangent cylinder.
The positive extrema seem to be correlated with negative radial shear
(bottom row of \Fig{effects}) or appear close to the surface in the equatorial region.
With overall parity $\tilde{P}=1.00$, the equatorial symmetry of
$\gamma_r$ is close to perfect.

The latitudinal component $\gamma_{\theta}$ is antisymmetric with
respect to the equator ($\tilde{P}=-0.99$).
It is negative (positive) 
in the northern (southern) hemisphere in the
lower half of the convection zone, with extrema close to the
tangent cylinder, while in the upper half of the convection zone
its sign is mostly opposite.
This means that turbulent pumping is poleward in the lower
half of the convection zone and equatorward in the upper half, as also
indicated by the streamlines of $\ggamma_{\rm pol}=\gamma_r \eee_r +
\gamma_\theta \eee_\theta$ in \Fig{turb}.
Interestingly, the pumping is strongly equatorward
in the same region where equatorward migration of the 
toroidal field is caused by the negative shear; see
Figs.~\ref{alpij}, \ref{effects} and \ref{turb} and compare with
Figs.~3 and 4 of \cite{WKKB14}.

The azimuthal component $\gamma_\phi$ is close to being equatorially
symmetric ($\tilde{P}=0.99$).
It is positive near the surface at low to mid latitudes 
as well as near the tangent cylinder at low latitudes, otherwise negative 
with minima just inside the latter
(see \Fig{effvel} for $\gamma_\phi$ in physical units).
All components show strong similarity with the turbulent pumping
coefficients presented in \cite{SRSRC07,SPD12}; only in the upper half
of the convection zone and in particular near the surface do we
find a different behavior.

To understand the effects of the off-diagonal components of $\aalpha$,
\cite{K91}, \cite{OSBR02} and, for example,
\cite{KKOS06} have advocated a view of component-wise pumping in
which three different pumping velocities $\ggamma^{(i)}$, acting on the
Cartesian component fields\footnote{No summation
over $i$ is applied here.} $\meanBB^{(i)}=\meanB_i \ee_i$,
are identified.
Under the condition that the latter are all solenoidal and the former are spatially constant,
it can be shown that each component field is advected as
$\partial_t \meanBB^{(i)} = -\ggamma^{(i)} \cdot\nab \meanBB^{(i)}$.
In the present context, however, neither of the stated conditions is satisfied.
Given that the poloidal and toroidal constituents of $\meanBB$
are solenoidal, we consider their
evolution separately, focussing on the terms related to turbulent pumping and mean velocities
\begin{align}
  {\partial_t \meanBBp} &=  \nab \times \left[
    \left(\meanUU_{\rm pol}+\ggamma_{\rm pol}\right) \times \meanBBp \right]+\dots\\
  {\partial_t \meanBBt}  &=  \nab \times \left[
    \left(\meanUU_{\rm pol}+\ggamma_{\rm pol}\right) \times \meanBBt +
    \left(\meanUU_{\rm tor}+\ggamma_{\rm tor}\right) \times \meanBBp
  \right]  +\dots \,. \nonumber
\end{align}
Thus, in the absence of all other effects, both $\meanBBp$ and
$\meanBBt$ are frozen into (but not advected by) the ``effective''
mean poloidal velocity
$\meanUU{}_{\rm pol}^{\rm eff}=\meanUUp+\ggamma_{\rm pol}$, while the
toroidal field is, in addition, subject to the source term
$\nab\times\big(\meanUU{}_{\rm tor}^{\rm eff} \times \meanBBp\big)$,
$\meanUU{}_{\rm tor}^{\rm eff}=\meanUUt+\ggamma_{\rm tor}$, representing
the winding-up of the poloidal field by the simultaneous effect of
differential rotation and toroidal pumping $\ggamma_{\rm
  tor}=\gamma_\phi\eee_\phi$.

We show the temporally averaged effective mean
velocities in comparison to $\meanUU$ alone in \Fig{effvel}.
For $\meanUU{}_{\rm pol}^{\rm eff}
=\meanU{}_r^{\rm eff} \eee_r + \meanU{}_\theta^{\rm eff} \eee_\theta$ (upper row),
turbulent pumping has a significant impact:
at high (low) latitudes its radial component  is dominated by the
strong downward (upward)
pumping such that there, $\meanU{}_{r}^{\rm eff}\approx 4 \meanU_{r}$,
while at the tangent cylinder, $\meanU{}_{\theta}^{\rm eff} \approx 2 \meanU_{\theta}$,
and the equatorward flow in the upper half of the
convection zone is also significantly enhanced.
Close to the surface the effective velocity has a strong equatorward
component. 
As a consequence, the whole meridional circulation pattern, as shown
by the streamlines in \Fig{effvel} is changed:
The three meridional flow cells aligned with the rotation vector
outside the tangent cylinder are no longer present in
$\meanUU{}^{\rm eff}_{\rm pol}$.
We note that, while at least $\brac{\overline{\rho \uu}}_t$ is solenoidal, 
no such constraint applies to $\ggamma_{\rm pol}$
and, hence, neither to $\meanUU{}^{\rm eff}_{\rm pol}$.
Near-surface patches of poloidal flux may, in principle, be able to
reach the bottom of the convection zone when transported by the
meridional circulation $\meanUUp$, albeit on a rather
involved route. 
However, this can hardly be accomplished by the effective
meridional circulation $\meanUU{}^{\rm eff}_{\rm pol}$ mainly due to its 
massive deviations from solenoidality.
Consequently, the flux transport dynamo paradigm seems to be
inconsistent with the presented simulations.
Even if helioseismic inversion were to determine
accurately the meridional circulation inside the solar convection zone,
the effective meridional velocity would still be unknown, because one
cannot measure $\ggamma$ inside the Sun.

The azimuthal flow $\mean{U}_\phi$ and hence the differential
rotation is only marginally modified by $\gamma_{\phi}$
(see \Fig{effvel}, bottom row).
However, at the surface it affects the radial shear
significantly, as shown in \Fig{NSsh}, where we plot the radial
derivatives of the rotation rate $\Omega$ and its effective counterpart
$\Omega^{\rm eff}=\meanU{}^{\rm eff}_\phi/r \sin\theta+\Omega_0$.
At low latitudes, the effective radial derivative becomes negative
whereas at mid latitudes it is weakly enhanced.
We note that simulations of the type employed here do not produce a
negative radial derivative as  found in the Sun \citep{KMCWB13,WKKB15}
where near the surface $\partial\ln\Omega/\partial\ln r=-1$
\citep[e.g.,][]{BSG14}  being possibly  responsible for the equatorward
migration of the toroidal field \citep[e.g.,][]{B05}.
Also, at this location the toroidal turbulent pumping can 
modify the effective radial shear and thus the magnetic field
generation.
\cite{SPD12} concluded that their radial and latitudinal turbulent pumping show a
strong influence on the magnetic field generation. This effect
was reported to become stronger for faster rotation.
However, the authors did not consider the effect of azimuthal
turbulent pumping, which can modify the $\Omega$ effect.

\begin{figure}[t!]
\begin{center}
\includegraphics[width=\columnwidth]{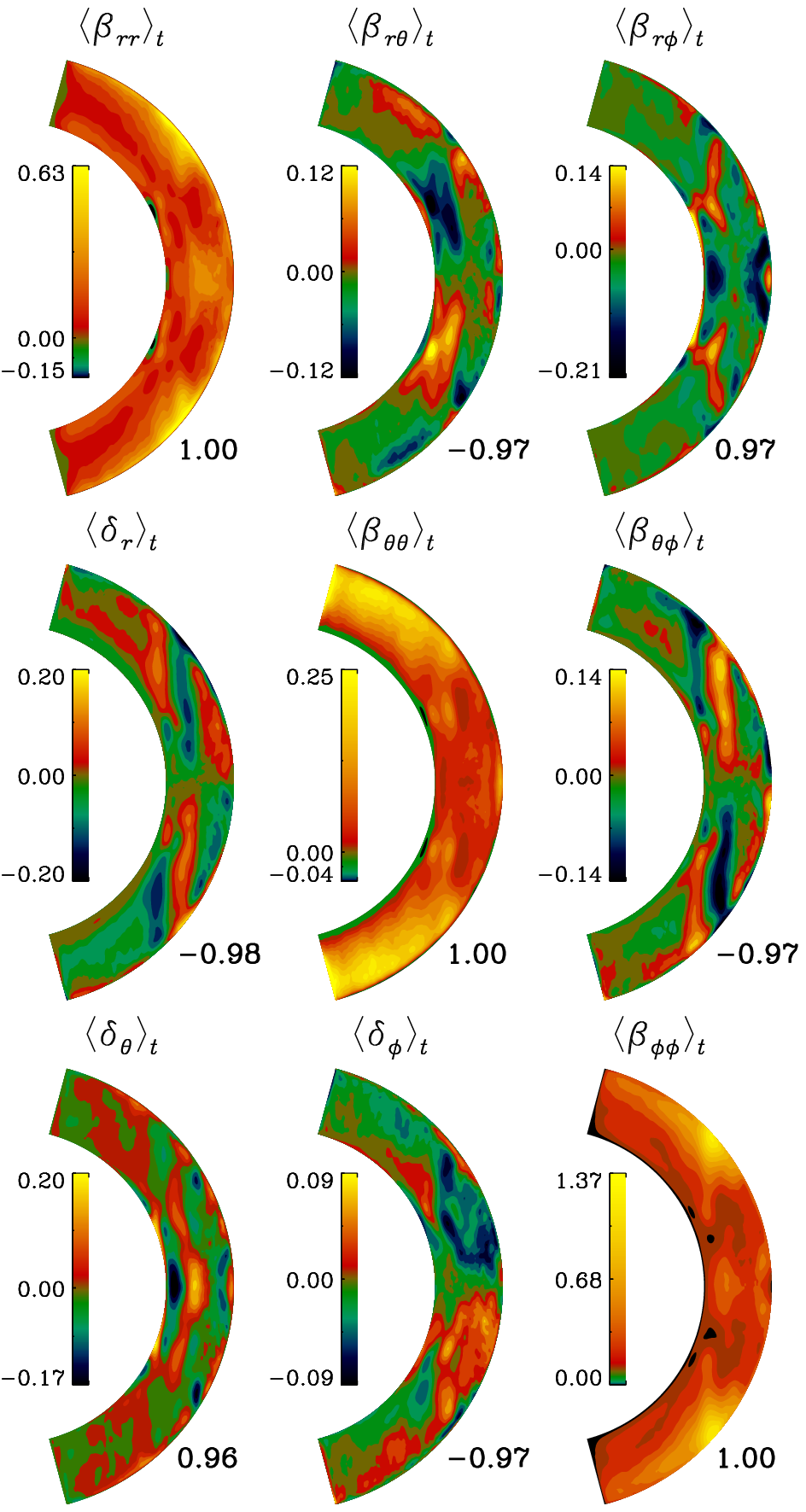}
\end{center}\caption[]{
Time-averaged components of $\bbeta$ and $\ddelta$,
normalized by $\etatz=\tau u^{\prime\,2}_{\rm
  rms}/3$.
Numerals on the {\it bottom right} of {\it each panel}: 
overall parity
$\tilde{P}$,  see \eq{eq:parity}.}\label{betij}
\end{figure}

\begin{figure}[t!]
\sidecaption{
\includegraphics[width=0.35\columnwidth]{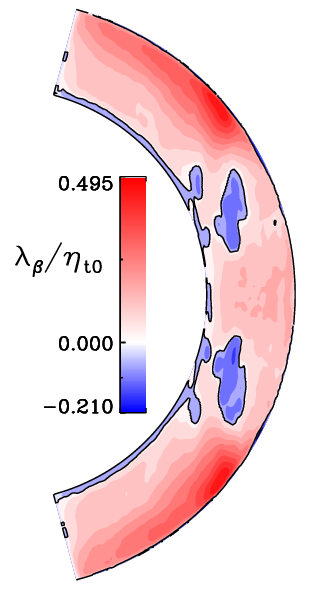}}
\caption[]{
Third root of the product of the three eigenvalues
of the time-averaged turbulent diffusivity tensor $\bbeta$,
$\lambda_\beta=\sqrt[3]{\beta_0^1\,\beta_0^2\,\beta_0^3}$,
normalized by $\etatz$.
Solid lines: zero level.
}\label{bet_pos}
\end{figure}

\begin{figure*}[t!]
\begin{center}
\includegraphics[width=15cm]{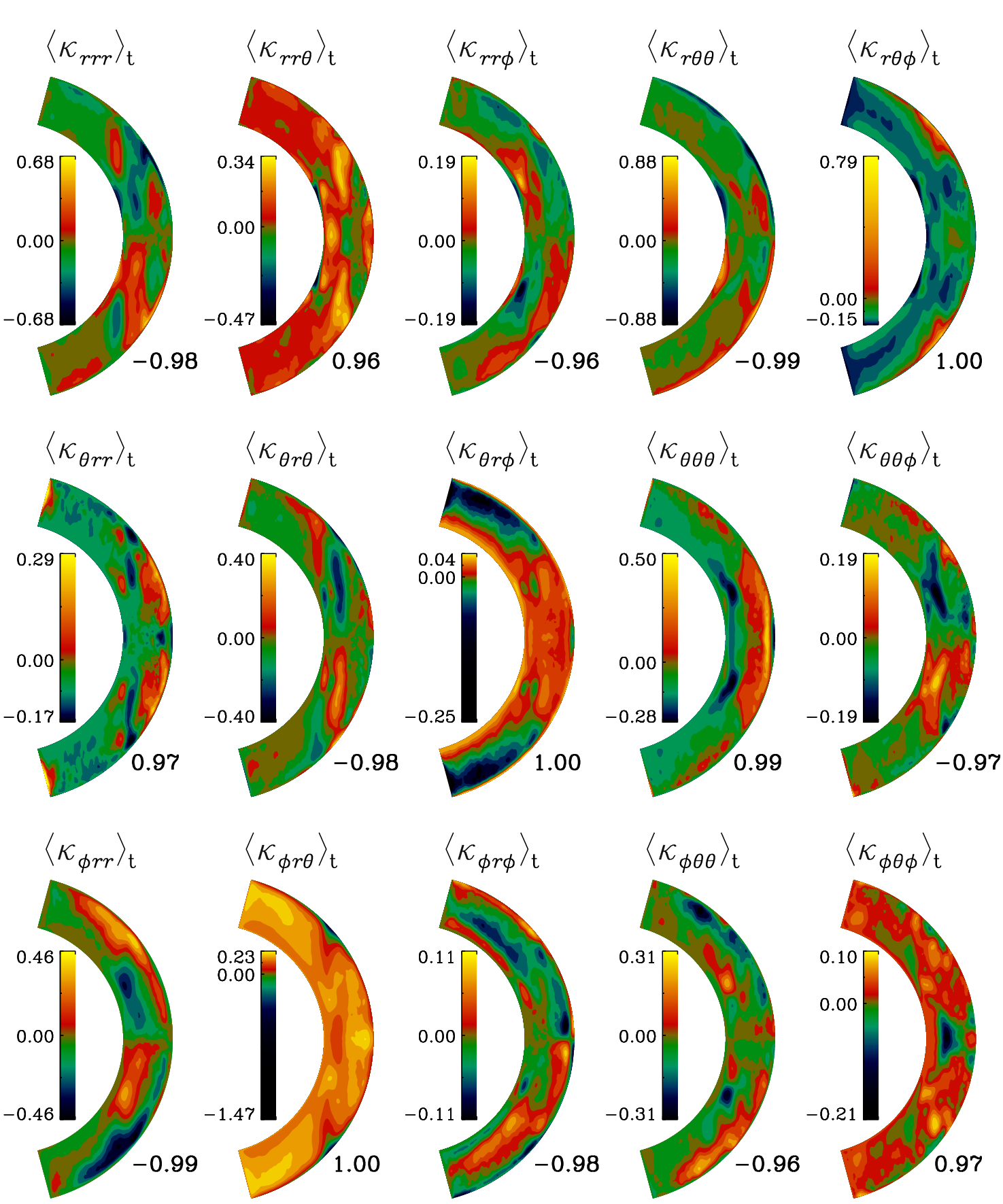}
\end{center}
\caption[]{
Time-averaged components of $\kkappa$,
normalized by $\etatz=\tau u^{\prime\,2}_{\rm rms}/3$.
Symmetric and irrelevant
components are not shown.
Numerals on the {\it bottom right} of {\it each panel:} volume averaged parity
$\tilde{P}$, see \eq{eq:parity}.
}\label{kapijk}
\end{figure*}

\subsection{Turbulent diffusion $\beta$, $\delta$ effect, and $\kappa$ term}
\label{sec:diff}

In \Fig{betij} we plot all components of the time-averaged $\bbeta$
and $\ddelta$ tensors.
All diagonal components of $\bbeta$ are mostly positive.
Overall, the values of $\bbeta$ are significantly smaller than the
turbulent diffusivity estimate $\etatz$ with the exception of $\beta_{\phi\phi}$
reaching 1.37$\etatz$ at the tangent cylinder near the surface.
These regions are also pronounced in $\beta_{rr}$, but its values are only
around half as large.
 With values
still five times smaller than $\beta_{\phi\phi}$, $\beta_{\theta\theta}$ is strongest at high latitudes.
The apparent dominance of $\beta_{\phi\phi}$ amongst the diagonal components
is partly an artefact of the ambiguity in defining the tensors in
\eq{eq:emf2} based on those in \eq{eq:EMFmu};
see \cite{SRSRC05}.
Effectively, $\beta_{rr}$ and $\beta_{\theta\theta}$ should be
multiplied by a factor of two for comparison with $\beta_{\phi\phi}$;
see the discussion of $\kkappa$ at the end of this section.
The overall parities of the diagonal components 
approach the ideal
value ($\tilde{P} = 1$), while those of the off-diagonal
components are still close 
($0.97\le|\tilde{P}| \le0.98$) 
with $\beta_{r\theta}$ and $\beta_{\theta\phi}$ being close to
antisymmetric and $\beta_{r\phi}$ close to symmetric.
In the northern hemisphere,
$\beta_{r\theta}$ is negative in the lower half of the convection zone and positive in the
upper half.
The contours of $\beta_{\theta\phi}$ are approximately aligned with
the rotation axis, showing a clear sign reversal outside the tangent cylinder
such that the sign of $\beta_{\theta\phi}$ is anti-correlated with that of
the latitudinal shear; see bottom row of \Fig{effects}.
$\beta_{r\phi}$ is particularly strong 
(positive) only near the equator.
In the geodynamo model of \cite{SRSRC07},
the diagonal $\bbeta$ components show a strong alignment and  
concentration near the tangent cylinders,
while in our work the components are most prominent
at high latitudes and
around the local minimum of $\Omega$.
Of the non-diagonal components, ($\beta_{\theta\phi}$) is similar outside
the tangent cylinder, ($\beta_{r\theta}$) is similar in the lower half
of the convection zone, and ($\beta_{r\phi}$) shows the opposite sign
with a similar pattern.

For inspecting the energetics of the system, it is interesting to study
whether or not $\bbeta$ is positive definite, that is, whether or not the effect of the
term $\nab \times \bbeta\cdot \nab\times\meanBB$ onto $\meanBB$ is
exclusively dissipative.  
We note that such a study was not performed in previous works
\citep{SRSRC07,SPD11,SPD12}.
To this end, we calculated the eigenvalues
$\beta_0^{1,2,3}$ of $\bbeta$ and depict 
the third root of their product in \Fig{bet_pos}. It is predominantly
positive,  adopting negative
values only at lower latitudes, and with less than half the moduli
compared to the maximum.
We note that this finding is not contradictory to any basic
principle as negative turbulent diffusivities have been conclusively
demonstrated to exist, albeit in laminar model flows
only \citep{DBM13}.
Comparison with the field generators in
\Fig{effects} suggests that a
possible generative effect from the negative definite $\bbeta$ is most
likely weak.

The coefficient $\ddelta$ is known to parameterize the ``$\OO \times
\JJ$ effect" appearing already in homogeneous anisotropic non-helical
turbulence \citep{KHR69,raedler76} when the preferred direction is
given by that of the global rotation. $\ddelta$ may also contain a
contribution from the ``shear-current effect'' which occurs already in
homogeneous non-helical turbulence under the influence of large-scale
shear \citep[see, e.g.,][]{RK03,RK04}.
We note that the major physical difference between turbulent diffusion
and the $\ddelta$ effect is that the latter does not lead to a change
in  mean field energy as $\meanJJ\cdot(\ddelta\times\meanJJ)=0$.

The coefficient $\ddelta$ shows no preferred sign for any of its
components; $\delta_r$ has the same pattern as $\beta_{\theta\phi}$
(contours aligned with the rotation axis, sign reversal), but opposite
values.
$\delta_\theta$ is strongest close to the equator with negative
values near the bottom and the upper part of the convection zone and
positive values in between and near the surface.
$\delta_r$ and $\delta_\theta$ are around two times larger than
$\delta_\phi<0.1\etatz$,
which is mostly negative (positive) in the northern (southern)
hemisphere outside the tangent cylinder and positive (negative) inside.
The overall parities of 
$\ddelta$ ($0.96\le|\tilde{P}| \le0.97$) 
are comparable with those of the off-diagonal
components of $\bbeta$  with
$\delta_\theta$ being symmetric and $\delta_r$ and $\delta_\phi$
antisymmetric.
We find that the contribution of the $\delta$ effect on the generation
and amplification of the magnetic field is small compared to the other
terms discussed in \Sec{sec:mag_gen}.
Similarly as for $\bbeta$, all $\ddelta$ components are consistent with
previous studies \citep{SRSRC07}; $\delta_r$ is similar outside the
tangent cylinder, while $\delta_\theta$ and $\delta_\phi$ are only
similar in the lower half of the convection zone.

The components of the rank-three tensor $\kkappa$ can be reduced from
27 to 15 independent components\footnote{By adopting symmetry in the
last two indices and dropping $\kappa_{i\phi\phi}$.};
see \Fig{kapijk} for the results of the current simulation.
All components are below $\etatz$ with the most dominant ones being
$\kappa_{rrr}$, $\kappa_{rr\theta}$, $\kappa_{r\theta\theta}$,
$\kappa_{r\theta\phi}$, $\kappa_{\theta\theta\theta}$.
Several of the profiles show alignment with the rotation axis.
The parity of all components of $\kkappa$ ($0.96\le|\tilde{P}|
\le1.00$) is similar to those of the off-diagonal components of
$\bbeta$.
We note here that considerable dissipative effects are ``hidden'' in the
$\kappa$ term. This can be seen by exploiting the freedom in defining
the components $b_{\mu\nu\phi}$; see Eq.~\eqref{eq:emf1}: They can be
chosen such that $\beta_{rr}$ and $\beta_{\theta\theta}$ adopt twice
their values. One can even go further and employ a choice by which all
off-diagonal components of $\bbeta$ disappear. The corresponding
value of $\lambda_\beta$ has then a maximum; of 1.7 instead of 0.495
(cf. Fig.~\ref{bet_pos}), indicating that the turbulent diffusion is
actually stronger by a factor $\approx 3.4$. 
Of course, changes in the choice of the $b_{\mu\nu\phi}$ influence, in general, all transport coefficients.

\begin{figure}[t!]
\begin{center}
\includegraphics[width=\columnwidth]{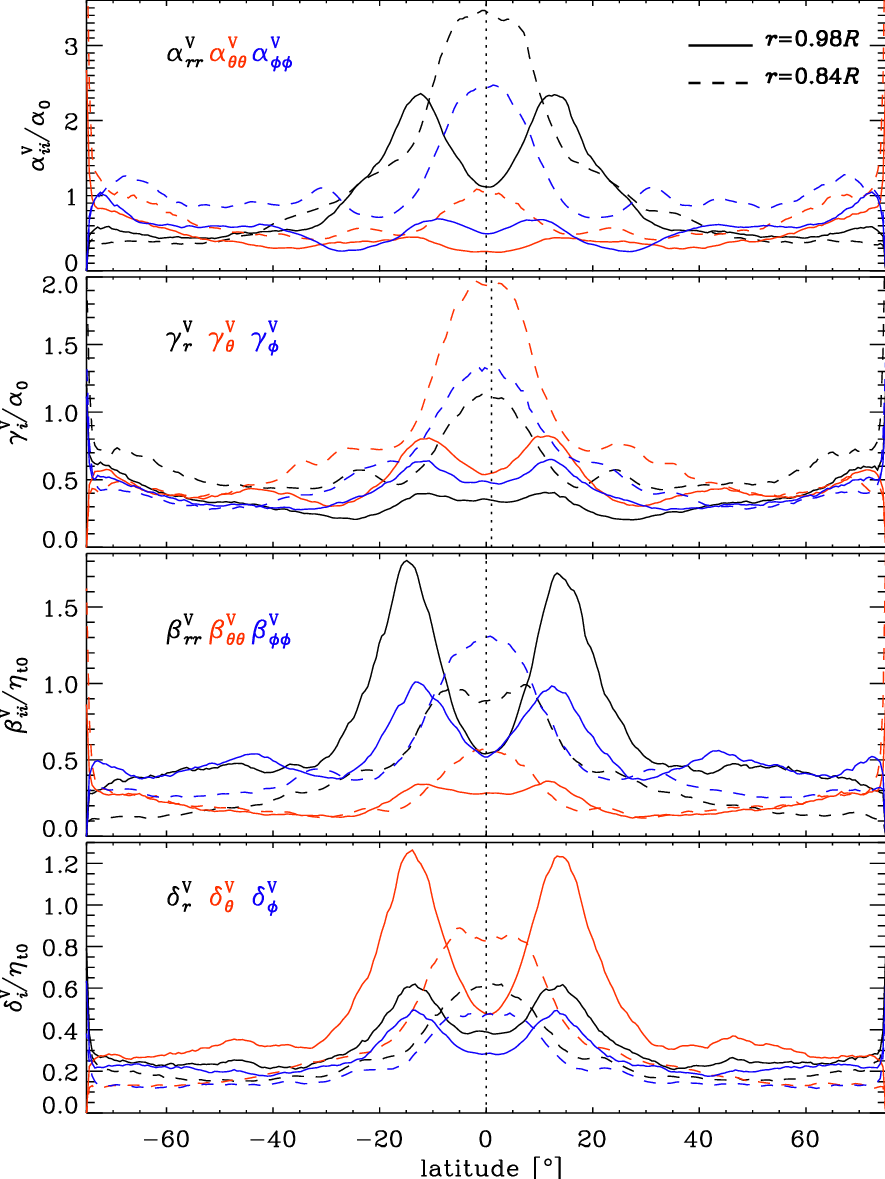}
\end{center}\caption[]{
Latitudinal distribution of 
the rms values 
$\alpha^{\rm V}_{ii}$, $\gamma^{\rm V}_{i}$, $\beta^{\rm V}_{ii}$ and
$\delta^{\rm V}_{i}$ 
of the transport coefficient variations
at $r=0.98\,R$ (solid) and
$r=0.84\,R$ (dashed).
}\label{var}
\end{figure}

\subsection{Variations in time}
\label{sec:var}

The transport coefficients show a strong variability in time and can hence
be divided into a constant (= time-averaged) part
and a part  with temporal average zero, called variation
(indicated by superscript $\rm v$),
such as 
\begin{equation}
\aalpha=\brac{\aalpha}_t+\vari{\aalpha}.
\end{equation}
In \Fig{var}, we plot
the rms values of the variations,
defined as
\begin{equation}
\alpha^{\rm V}_{ij}=\sqrt{\bra{{\alpha^{\rm v}}^2_{\!\!\!ij}}_t},\;
\gamma^{\rm V}_{i}=\sqrt{\bra{{\gamma^{\rm v}}^2_{\!\!\!i}}_t},\;
\beta^{\rm V}_{ij}=\sqrt{\bra{{\beta^{\rm v}}^2_{\!\!\!ij}}_t},\;
\delta^{\rm V}_{i}=\sqrt{\bra{{\delta^{\rm v}}^2_{\!\!\!i}}_t}
 \label{eq:var_rms}
\end{equation}
(we emphasize the capital V in the symbol).
For all shown coefficients these are stronger at lower-   than at
higher latitudes.
Near the surface ($r=0.98$) the variations have their maxima around
$\pm (10\ldots15)^\circ$
latitude.
This distribution indicates a strong influence of the mentioned
poleward migrating high-frequency constituent.
However, in the middle of the convection zone
($r=0.84$) the maxima are around the equator, where the high-frequency
constituent is not present \citep{KKOBWKP15}.
In addition, at mid to high latitudes, the variations also show
significant values.
The variations of $\alpha_{ii}$ and $\gamma_i$ are
roughly equal to their time averages
near the surface, but significantly bigger in the middle of the
convection zone near the equator. 
Furthermore, $\beta^{\rm V}_{rr}$ and $\beta^{\rm V}_{\theta\theta}$
are significantly stronger than $\brac{\beta_{rr}}_t$ and
$\brac{\beta_{\theta\theta}}_t$, respectively, but 
$\beta^{\rm V}_{\phi\phi}$ is only about one half of
$\brac{\beta_{\phi\phi}}_t$.
The variations of the $\delta_i$ also exceed  their
time-averages by several times.
In his PhD thesis, \cite{Schrinner2005} also presented the standard
deviations of the transport coefficients for a time-dependent dynamo.
There, he found values of a relative strength of 0.4 to 0.7 for the
diagonal $\aalpha$ components, which is somewhat lower than what we find.
However, the convection in his simulation was not as vigorous
as in ours.
Thus, we can conclude that the time variation of the coefficients 
may play an important role in the evolution of the magnetic field.
Further analysis shows that the variation distributions can be modeled
by Lorentzian profiles and are therefore consistent with a random
process.
However, the variations also have non-random contributions, 
which show tight relations to the mean magnetic field evolution.
This is discussed further in \Sec{sec:cycl}.

\begin{figure}[t!]
\begin{center}
\hspace*{-0mm}\includegraphics[width=0.95\columnwidth]{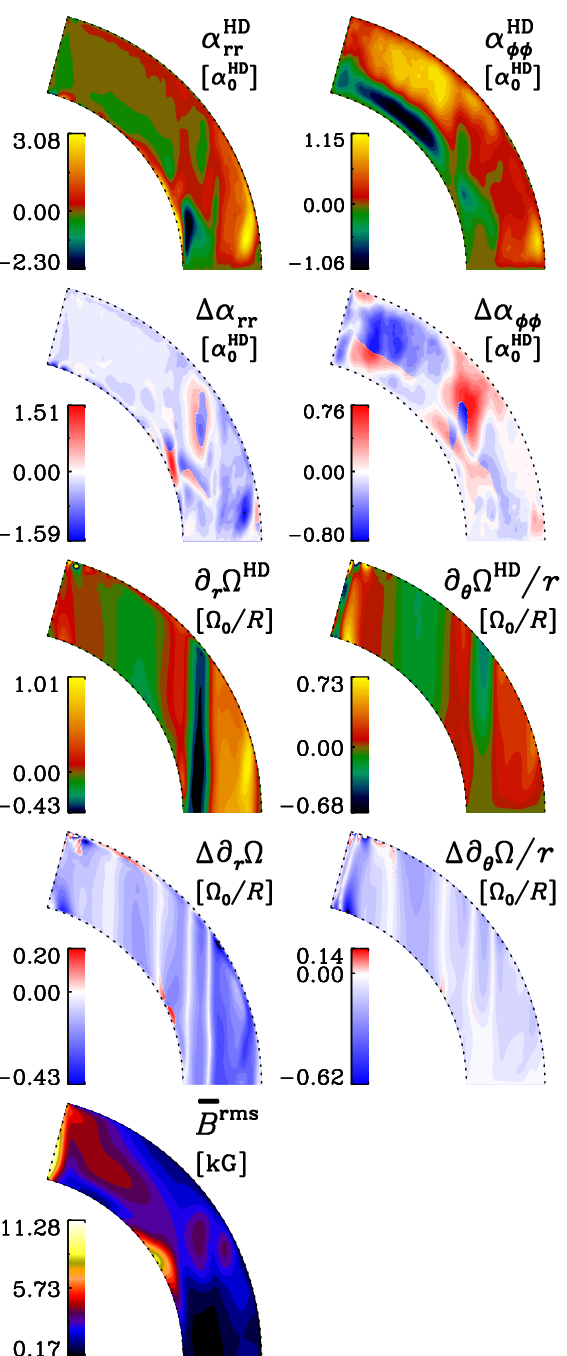} 
\end{center}
\caption[]{
Magnetic quenching of $\alpha$ effect and shear.
{\it From top to bottom:}\\[.5mm]
$\alpha_{rr,\phi\phi}^{\rm HD}$ from HD,
$\Delta\alpha_{rr,\phi\phi}$ from MHD and HD, radial and latitudinal shear 
$\partial_r \Omega^{\rm HD}$, $\partial_\theta \Omega^{\rm HD}/r$ from HD,
$\Delta \partial_r \Omega$, $\Delta \partial_\theta \Omega/r$ from MHD and HD,
$\meanB^{\rm rms}$ (rms over full time series).
The $\Delta$ quantities contain
the sign of the corresponding HD quantities to highlight enhancement
(red) and quenching (blue).
}\label{quench}
\end{figure}

\begin{figure*}[t!]
\begin{center}
\includegraphics[width=2.\columnwidth]{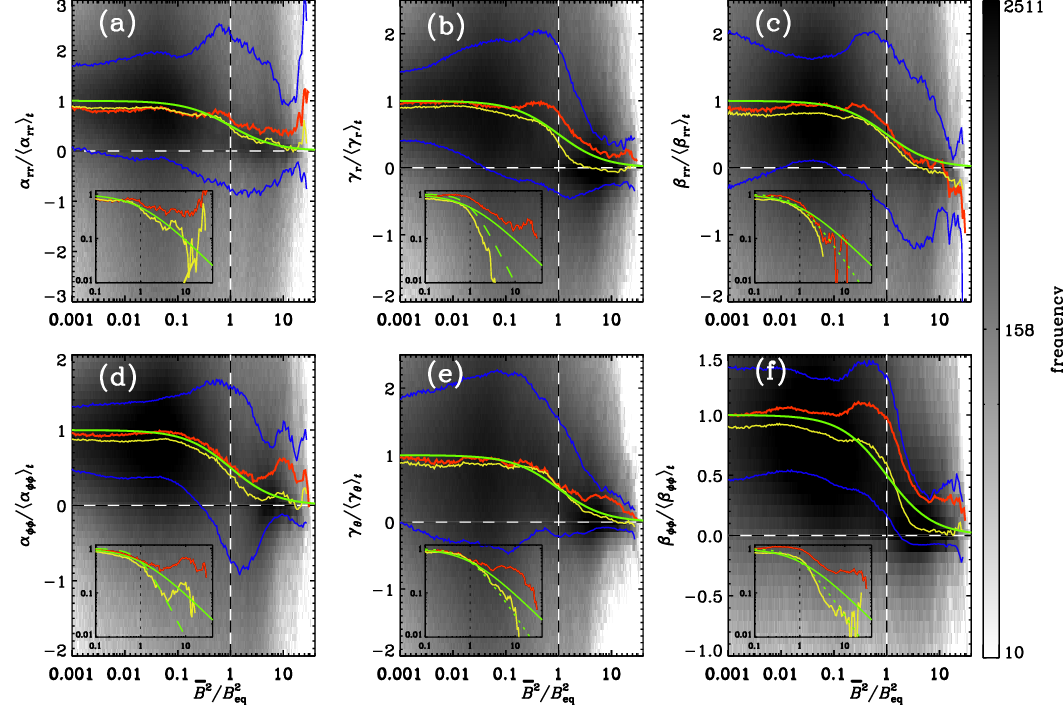}
\end{center}\caption[]{
Quenching of transport coefficients shown as 2D histograms of
$\alpharr$ (a), $\gamma_r$ (b), $\beta_{rr}$ (c), $\alphapp$ (d),
$\gamma_\theta$ (e), $\beta_{\phi\phi}$ (f), normalized by their time
averages, over the normalized 
energy density of the mean field
$\mean{B}{}^2/\Beq^2$.
Data are taken from
the entire domain and the
whole saturated stage.
Red and yellow lines: mean and median, 
respectively;
blue contours: margins of range, in which 50\% of the values lay.
Green line: 
 $\sim 1/\big(1+\mean{B}\,{}^2/\Beq^2\big)$.
Inlays: average and median
in log-log scale for
zoomed-in range; dotted and dashed green:
$\sim 1/\big(1+\mean{B}\,{}^3/\Beq^3\big)$
and
$\sim 1/\big(1+\mean{B}\,{}^4/\Beq^4\big)$,
respectively.
}\label{quench2}
\end{figure*}

\subsection{Magnetic quenching}
\label{sec:mag_quen}

To investigate the alteration of the
transport coefficients
by the mean field, we first compare the MHD run
to the corresponding HD run.
In \Fig{quench}, we show the absolute differences of the time-averaged
dominant diagonal components $\alpharr$ and $\alphapp$,
employing the quantity
$\Delta\alpha_{ij}=\big(\alpha^{\rm MHD}_{ij}-\alpha^{\rm
  HD}_{ij}\big)\times\sgn \alpha^{\rm HD}_{ij}$.
Here, multiplying with the sign allows 
to distinguish between enhancement (red) and quenching (blue).
$\alpharr$ is mostly quenched, in particular at low latitudes.
$\alphapp$ shows enhancement in the region of
strong negative latitudinal shear, and also in the
lower half of the convection zone at high latitudes.
Strong quenching of $\alphapp$ occurs mostly at high latitudes, where
the rms value of the mean field is particularly
strong; see the last row of \Fig{quench}.

The other quantities
also exhibit alterations due to the presence of the
magnetic field (not shown here).
The upward pumping $\gamma_r$ near the surface is much stronger
in the MHD case compared to the HD one
whereas at high latitudes $\gamma_r$ is suppressed.
$\gamma_\theta$ and $\gamma_\phi$ are also suppressed in the region of
strong negative radial shear and strong toroidal field at mid and high latitudes.
$\beta_{\phi\phi}$ and $\beta_{rr}$ are, in general, suppressed compared
to the HD case, in particular near the surface at mid-latitudes.
However, they exhibit enhancement in
the regions with strong toroidal field and negative shear at mid-latitudes
where $\beta_{r\phi}$ and $\beta_{\theta\phi}$ are suppressed 
while being enhanced everywhere else.
$\beta_{\theta\theta}$ turns out to be enhanced everywhere.
In comparison with the HD case, $\delta_r$ is suppressed
in the proximity of the tangent cylinder and at high latitudes.
Similarly to $\delta_\phi$, $\delta_\theta$ is suppressed near the surface at low latitudes and
in the region of negative shear.

As indicated by the quantities
$\Delta(\partial_{r,\theta}\,\Omega)$, defined analogously to
$\Delta\alpha_{ij}$,  the shear is also quenched,
similar to what is seen in, for example, \cite{BBT05}, \cite{BuSim06},
\cite{Aubert05}, \cite{GDW12} and in particular recently for models of
solar-like stars \citep[e.g.,][]{FF14,KKKBOP2015,KKOBWKP15,KKOWB17}.
This means that the quenching of the turbulence by the mean field results
in the modification of both the differential
rotation generators \citep{KKKBOP2015} and the mean $\EMF$.
Therefore, along with its direct influence via the fluctuating velocity,
the mean field also has a more indirect effect
on the transport coefficients via the mean flows.

Given the strong variability of the mean field both in space and time,
we suggest using the corresponding variability of the
transport coefficients for deriving their functional dependencies on
the mean field,  thus opening the gateway to predictive mean-field
models.
Unfortunately, we have to expect that these dependencies involve
complex and nonlocal relations not only with the mean field, but
also with its derivatives, for example, the mean current density.
Currently, no appropriate mathematical model and hence no numerical
scheme identifying these relations is available.
However, a starting point could be the model of \cite{RB12} with
$\meanBB$- and $\meanJJ$-dependent coefficients
in  the evolution equation for $\EMF$.
Here we restrict ourselves to establishing a rough relationship between the magnitudes
of the transport coefficients and the modulus of the mean field including all points in the domain
and all instants in time. 
Of course we cannot hope to find anything close to unique functional
dependencies.
Instead, for a certain field strength many different values of a transport coefficient
are found across the spatio-temporal domain. 
Hence, if at all, relationships can only be identified in a statistical sense. 

In \Fig{quench2}, we plot 2D histograms of some of the most important
coefficients depending on the energy density of the mean field,
normalized by its equipartition value 
$\Beq=\big(\mu_0\mean{\rho}\big)^{1/2}\urmsp$.
For this, the whole saturated stage (18 activity cycles) 
was considered, while smoothing the
coefficients over six points in space and time ($\approx$ a tenth of
an activity cycle).
We use 200 bins for both the normalized coefficients
(in the interval  $[-10,10]$) and the normalized field (in the
interval $[10^{-3},40]$.
Although the scatter is large, in most of the coefficients 
there is significant quenching detectable where
the mean field becomes dynamically important
($|\meanBB |\gtrsim \Beq $).
The average and median values for weak
mean field are slightly below the time-averaged quantities, but we
still use the latter as proxies for the unquenched
values.

\begin{figure*}[p]
\begin{center}
\includegraphics[width=\textwidth]{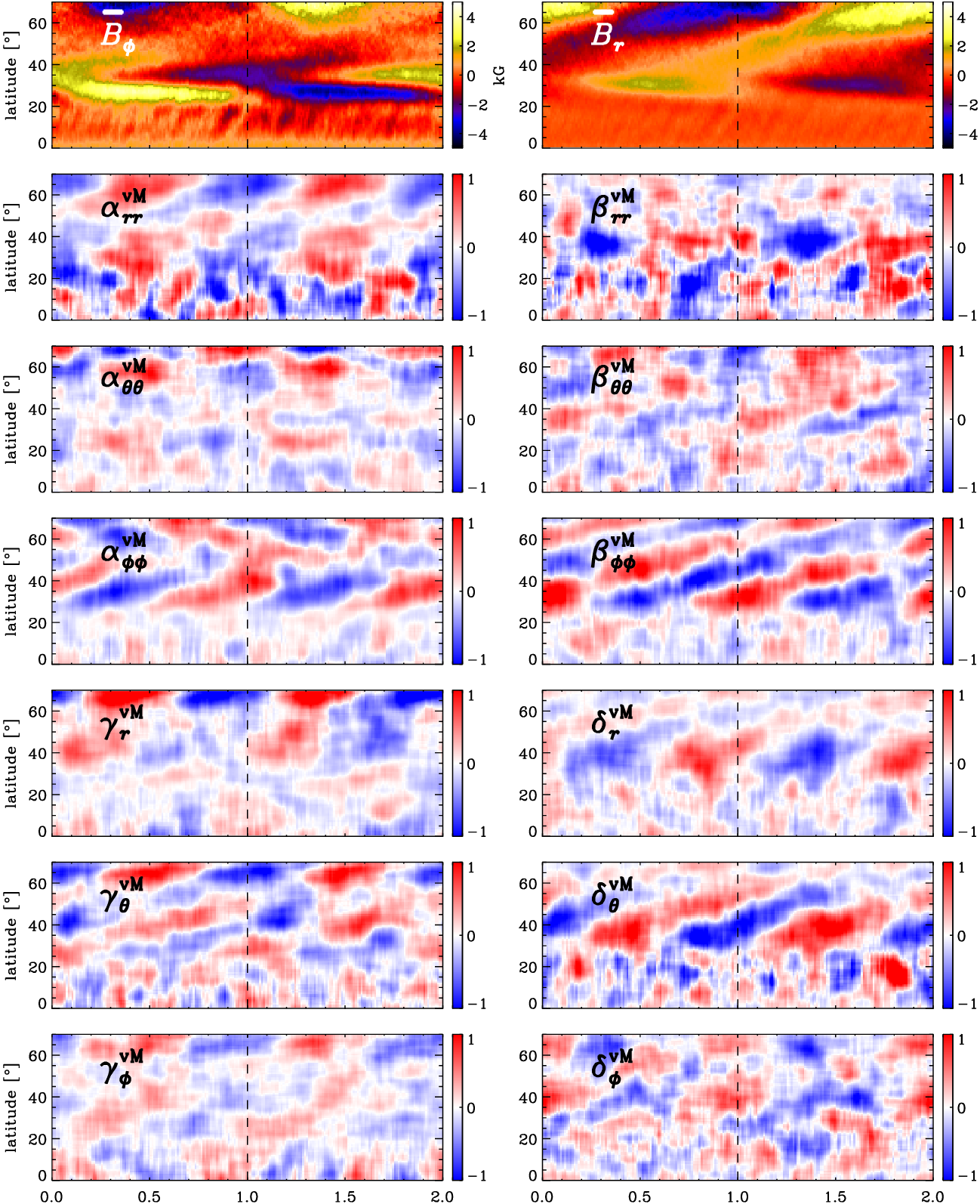}
\end{center}\caption[]{
Average cycle dependency of selected  
transport coefficients.
Mean azimuthal and radial magnetic field, $\mean{B}_{\phi,r}$ (top),
 together with the temporal variation of the diagonal components of
$\aalpha$ along with $\ggamma$ (left) as well as the diagonal
components of $\bbeta$ along with $\ddelta$ (right) near the surface
($r=0.98\,R$) on $\theta$-$t$ plane.
The data is obtained from a typical cycle; see
\Sec{sec:mag_gen}.
The coefficients are symmetrized according to their theoretical parity
for a perfectly equatorially symmetric flow.
The color scales are normalized to
highlight the patterns.
}\label{avcyl}
\end{figure*}

The coefficient $\alpharr$ in \Fig{quench2}(a) is quenched
for $|\meanBB|\gtrsim2\Beq$,
with the median following approximately a quadratic
characteristic $\sim 1/\big(1+\mean{B}{}^2/\Beq^2\big)$.
For stronger fields, median and mean
show inconclusive
behavior, including both enhancement and quenching,
and hence do not  follow any simple analytic 
dependency.
For $\alphapp$, \Fig{quench2}(d), the behavior is similar, 
albeit with a median closer to quartic behavior $\sim 1/\big(1+\mean{B}{}^4/\Beq^4\big)$.
$\gamma_r$, \Fig{quench2}(b), shows even enhancement
until $0.7 \Beq$, 
but then a decrease in mean and median with
a less-than-quadratic and stronger-than-quartic characteristic,
respectively.
The median of the
latitudinal pumping $\gamma_\theta$, \Fig{quench2}(e), follows
a cubic characteristic $\sim 1/\big(1+\mean{B}{}^3/\Beq^3\big)$
relatively closely, but with a much
shallower-than-quadratic mean.
For the turbulent diffusion, \Fig{quench2}(c) and (f), the
coefficients are enhanced below 
$\mean{B}=(0.4\ldots0.5)\Beq$
and decrease then with quadratic or even cubic decline
(in the medians).
To summarize, none of the coefficients follow exactly one single analytical quenching
formula.

\cite{SRSRC07} studied the quenching of the turbulent transport
coefficients not by comparing with a purely hydrodynamic run or 
employing a time-dependent magnetic field, but used a series of simulations with
increasing magnetic field strengths.
They found that $\alphapp$ and all diagonal $\bbeta$ components first
increase with growing magnetic field, but decrease when the 
Lorentz force becomes comparable to the Coriolis force.

\subsection{Cyclic variation}
\label{sec:cycl}

\begin{figure}[t!]
\begin{center}
\includegraphics[width=\columnwidth]{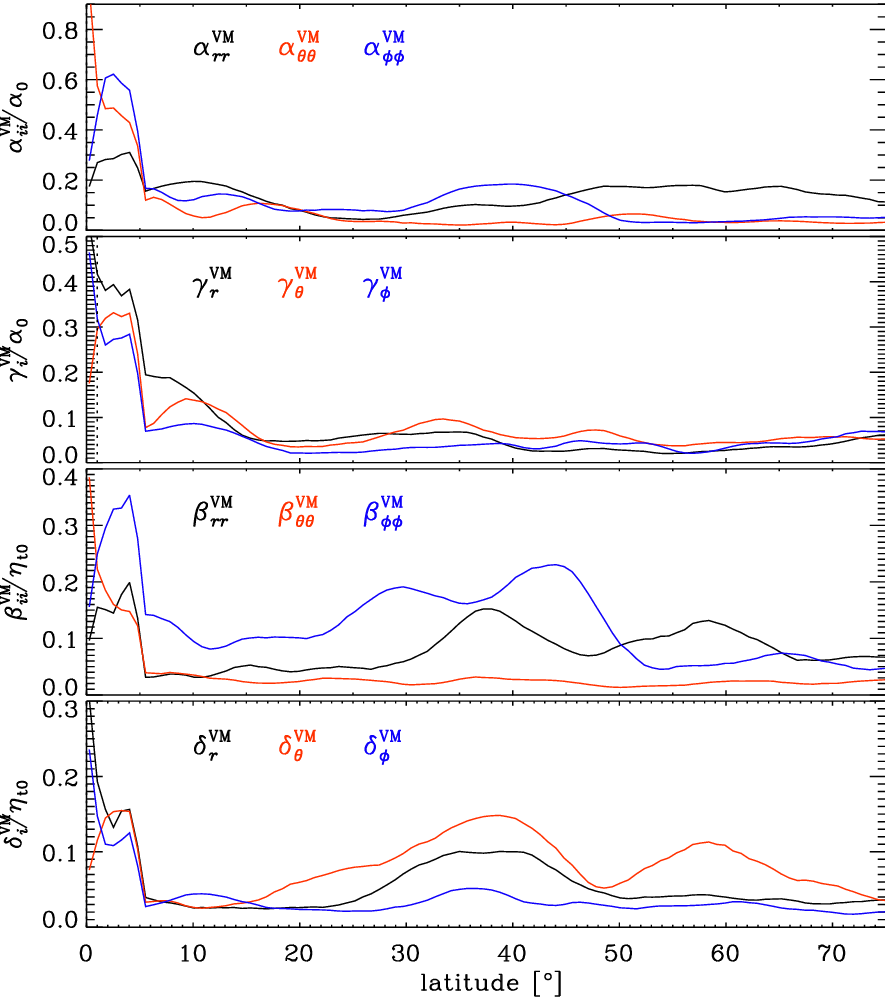}
\end{center}\caption[]{
Latitudinal distribution of 
the rms values 
$\alpha^{\rm VM}_{ii}$, $\gamma^{\rm VM}_{i}$, $\beta^{\rm VM}_{ii}$ and  
$\delta^{\rm VM}_{i}$ 
of the cyclic variations of the transport coefficients
at $r=0.98\,R$.
}\label{var_cyl}
\end{figure}

As a consequence of the cyclic variation of the mean field, 
the transport coefficients also vary cyclically.
In \Fig{avcyl}, we plot the cyclic variations of the diagonal
components of $\aalpha$ and $\bbeta$, along with $\ggamma$, $\ddelta$ and the
corresponding toroidal and radial mean field 
for a typical cycle, the definition and computation of which was
described in \Sec{sec:mag_gen}.
By folding the cycles on top of one another, one basically filters out
the variation based on the cyclic magnetic field.
To distinguish the cyclic variations
 from the time variations discussed in \Sec{sec:var}, we add
the superscript ``M'' to the symbols; for example,\ $\alpha^{\rm vM}_{ii}$.
All coefficients show clear cyclic variations
with the activity cycle period (= half the magnetic cycle period),
owed to the quadratic effect of the mean field on the velocity fluctuations.
In many cases, this is clearly visible
at high latitudes.
$\alphatt$
and $\delta_\phi$ seem to be predominantly quenched by the toroidal field,
while $\alpharr$, $\gamma_\phi$, $\gamma_r$, $\gamma_\theta$,$\delta_r$ and
the shown components of
$\bbeta$ seem to be predominantly quenched by the radial field as
indicated by the pattern at high latitudes.
At lower latitudes, the variations exhibit
time scales shorter than the activity cycle, which
cause a noisy signal when folding the cycles. 
The fast poleward migrating 
constituent of $\mean{B}_\phi$
near the surface at low latitudes
(ghosts of which are visible in \Fig{avcyl}),
discussed in \cite{WKKB14} and \cite{KKOBWKP15}, is one candidate for
causing such a signal.
As discussed in \cite{KKOBWKP15} for a similar run, this high-frequency 
dynamo mode is highly incoherent over time, that is, cycle length and phase change on short
time scales, which could well be the cause of the noisy appearance when
averaged over several cycles.

To quantify the variations further,  in
\Fig{var_cyl} we plot their rms values, defined analogously to Eq.~\eqref{eq:var_rms}.
For all shown coefficients, the cyclic variations are around four times
smaller than the non-cyclic variations; they are, however, still
comparable with the time-averaged values.
Also here the coefficients vary more strongly at lower than at higher
latitudes. 
The variation in the $\phi\phi$ components of $\aalpha$ and $\bbeta$
are stronger than in the other components.
Otherwise, the behavior and values of all components seem to be similar.
Given that the relative cyclic
variations in the mean flows are at most around 10$\%$ (see the
discussion in \cite{KKOBWKP15}),
we conclude 
that the back-reaction of the mean field onto the terms
generating it is predominately via the variation of the 
transport coefficients.

\subsection{Magnetic field propagation}
\label{sec:park}
\begin{figure}[t!]
\begin{center}
\includegraphics[width=\columnwidth]{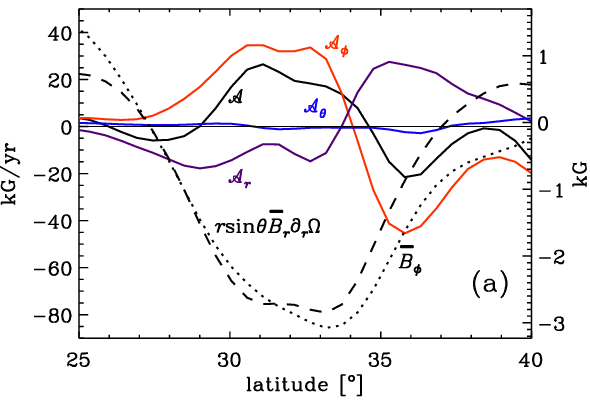}
\includegraphics[width=0.4\columnwidth]{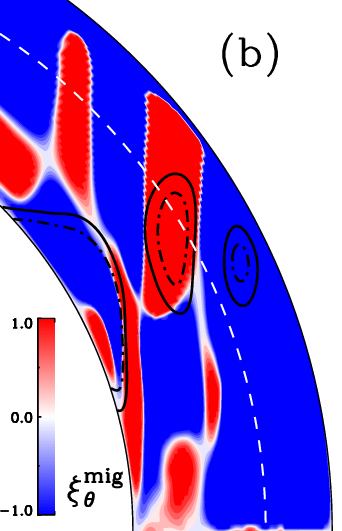}
\includegraphics[width=0.4\columnwidth]{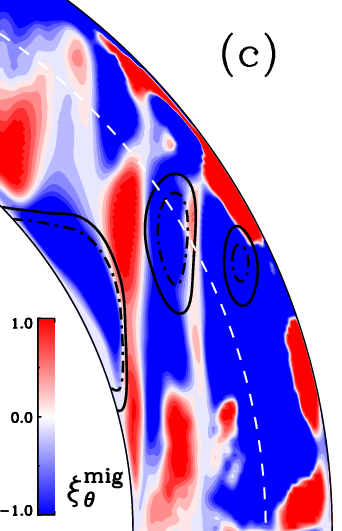}
\end{center}\caption[]{
(a) Radial $\alpha$ effect together with radial shear and
$\meanBBt$
at $r=0.9\,R$ between 25$^\circ$ and 40$^\circ$
latitude.
{\it Black solid:} ${\cal A}= \big(\nab\times\alpha_{ij}\mean{B_j \eee_i} \big)_r$.
\,{\it Red:} ${\cal A}_\phi= \big(\nab\times\alpha_{\phi \phi}\mean{B_\phi}\eee_\phi \big)_r$.
{\it Purple:} ${\cal A}_r= \big(\nab\times\alpha_{\phi r}\mean{B_r}\eee_\phi \big)_r$.
{\it Blue:} ${\cal A}_\theta=\big(\nab\times\alpha_{\phi \theta}\mean{B_\theta}\eee_\phi \big)_r$.
{\it Black dashed:} $r\sin\theta\,\mean{B_r}\partial_r\Omega$.
{\it Black dotted:} $\mean{B_\phi}$.\\
(b) and c): Latitudinal propagation 
direction $\xi^{\rm mig}_{\theta}$ of the mean 
field as predicted by the Parker-Yoshimura rule \eqref{Eq:eqm}, 
using $\alphapp$ (b) and $\alphapp^{\rm BS}$ (c, see \Sec{sec:bs}),
together
with $\Btt$ as black contours for 3 (solid) and 3.5 kG
  (broken).
The color scale is truncated between $-1$ and 1 to emphasize the sign
of $\xi^{\rm mig}_{\theta}$. The dashed white lines indicate $r=0.9\,R$.
}\label{alp_eff}
\end{figure}

As discussed in \cite{WKKB14}, the occurrence of the
equatorward-propagating magnetic field
found in \cite{KMB12} can be well explained by the Parker-Yoshimura rule
\citep{P55,Yos75} using $\alpha_K+\alpha_M$ as the relevant scalar $\alpha$.
For the rule to be applicable, the $\Omega$ effect must be dominant over
the toroidal $\alpha$ effect, and the poloidal $\alpha$ effect must
be expressible with a single (possibly position-dependent) scalar by
$\nab\times (\alpha \mean{B}_\phi\eee_\phi)$.
Having now all transport coefficients at hand allows us to investigate why
the Parker-Yoshimura rule provides such a good description.
To show this, we focus on the mid-latitude region where the shear is negative,
causing the generation of equatorward migrating toroidal field
$\meanBBt$.
There, the contribution $\sim\mean{B_r} \nabla_r \Omega$ to the
$\Omega$ effect dominates the generation of the toroidal field.
So the radial component
is the important part of the poloidal
field in the dynamo wave.
In \Fig{alp_eff}(a) we plot the 
contributions to the radial
$\alpha$ effect ${\cal A}$, named ${\cal A}_r$, ${\cal A}_\theta$, and ${\cal A}_\phi$
and defined in the caption.
The latitudinal $\alpha$ effect shows a similar behavior.
Obviously, the contribution related to $\alphapp$ (${\cal A}_\phi$,
red line) is
indeed dominant in the region where the toroidal field and
the negative shear are strong.
Consequently,
we now use $\alphapp$ to determine the equatorward
propagation 
direction:
\begin{equation}
\ssss^{\rm mig}(r,\theta) \sim -\alphapp \eee_\phi\times\nab\Omega.
\label{Eq:eqm}
\end{equation}
Figure~\ref{alp_eff}(b) shows that the result 
is consistent with the actual propagation direction.\footnote{The rule
  does not exclude dynamo waves propagating along directions inclined
  with regards to\ the isocontours of $\Omega$. The highest growth rate,
  however, occurs for aligned propagation. We note that in the saturated
  nonlinear stage, a kinematically subdominant mode may nevertheless be
  prevalent.}
Using $\alpha_K+\alpha_M$ instead of $\alphapp$ works for this run only
by chance as their signs are the same in the
region of interest.
However, in general, the Parker-Yoshimura rule using
$\alphapp$  will not always work as other components of $\aalpha$
may give more important contributions.
Parker dynamo waves have also been found in numerous Boussinesq
\citep[e.g.,][]{BuSim06,SPD12} and anelastic dynamo simulations
\citep[e.g.,][]{GDW12}.
In particular, in these studies it was found that the oscillation
frequency can be well described by a simplified dispersion relation of the
Parker dynamo wave.
To show that the magnetic field propagation in our simulation is consistent with a
Parker dynamo wave, we estimate its period, as described by
\cite{P55} and \cite{Yos75}, using a wave ansatz $\sim\exp(ik_\theta
r\theta - i \omega t)$ which leads to the frequency
\begin{equation}
\omega_{\rm
  PY}=\left|\frac{\alphapp\,k_\theta}{2}r\cos\theta\frac{\partial\Omega}{\partial  r}\right|^{1/2},   
\label{eq:wpy}
\end{equation}
where we only consider latitudinal propagation.
With $k_\theta r\approx1$, we obtain
periods of two to four years in the region
of interest, which is close to the actual period of around five years.
As investigated in detail by \cite{KKOBWKP15} and \cite{OKP16}
for a similar run, the
main dynamo mode, present in the upper and middle
part of the convection zone, shows a strongly variable cycle period
that remains coherent only over two to five cycles, with values of
between four and eight years.
The value of $k_\theta$ is a lower limit and a higher one
might be more reasonable, which would lead to a shorter period,
while still retaining the correct order of magnitude.
Taking into account the strong simplifications leading to \eq{eq:wpy},
for example the neglect of anisotropic contributions to $\alpha$,
the predicted period fits
the actual one rather well.

\subsection{Comparison with multidimensional regression method}
\label{sec:bs}
\begin{figure}[t!]
\begin{center}
\includegraphics[width=\columnwidth]{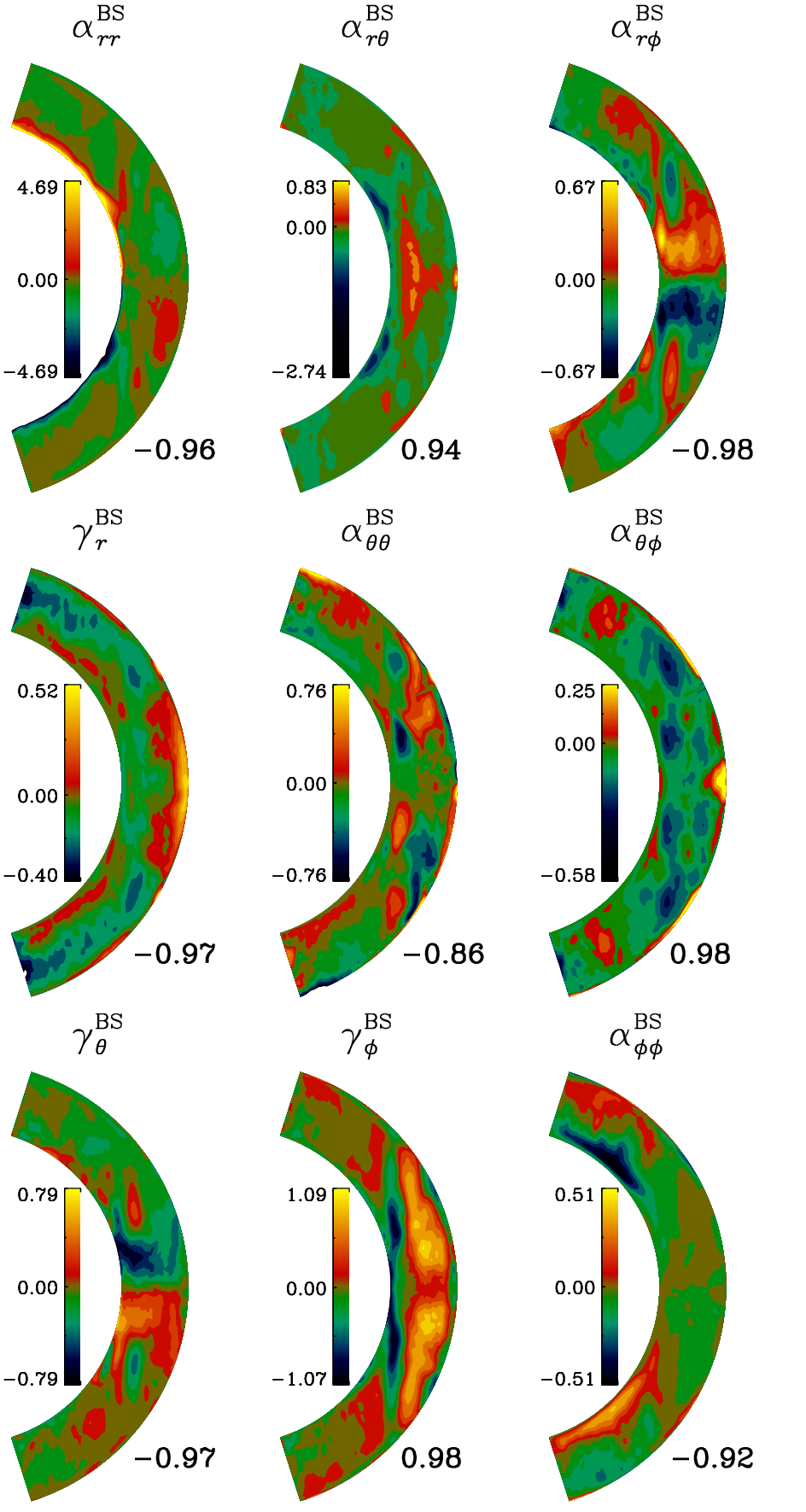}
\end{center}\caption[]{
$\aalpha$ and $\ggamma$ determined via the reduced
BS method as in \cite{RCGBS11};
see also 
\Fig{bsII}.
We note that in $\alpha_{rr}$ the extrema are actually ten times bigger than indicated 
in the color-bar.
Numerals  at the {\it bottom right} of {\it each panel}: overall parity $\tilde{P}$, see \eq{eq:parity}.
}\label{bs}
\end{figure}
\begin{figure*}[t!]
\begin{center}
\includegraphics[width=2.05\columnwidth]{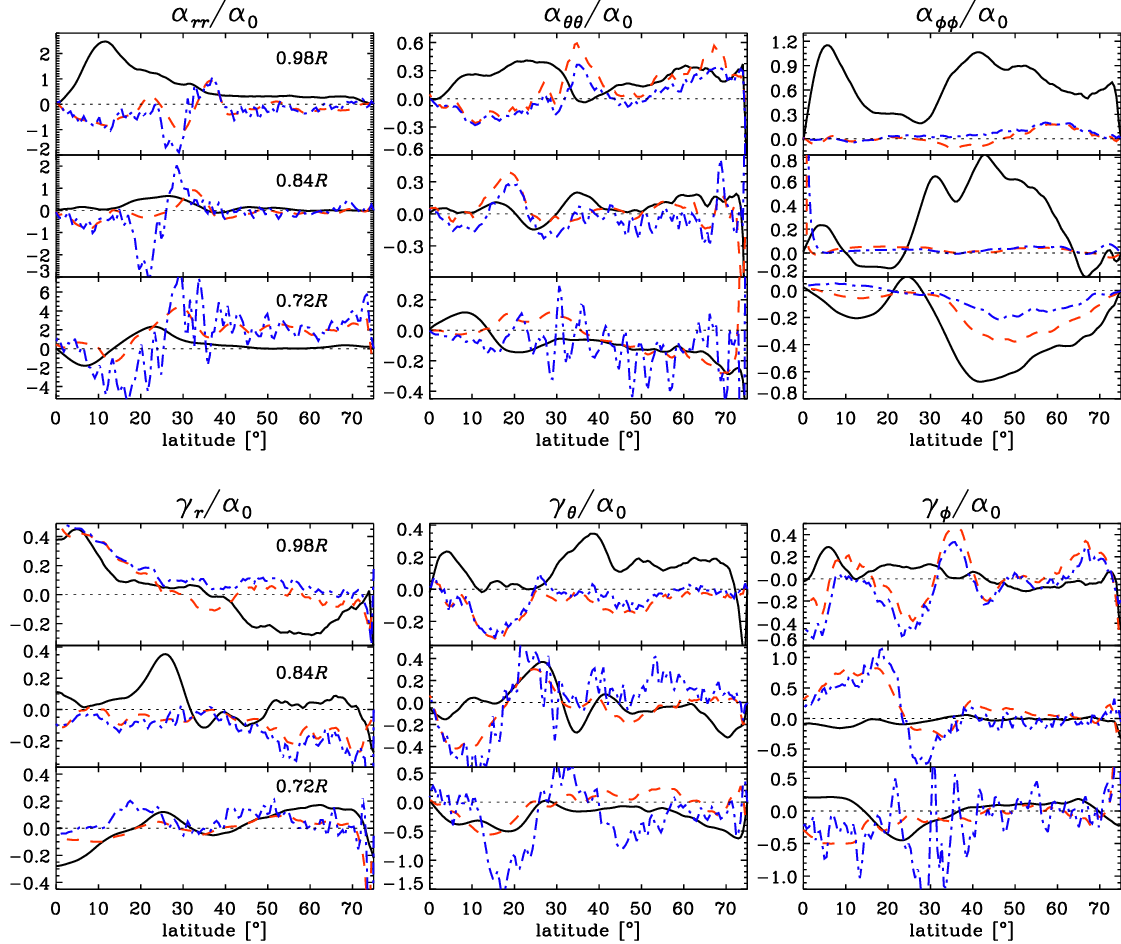}
\end{center}\caption[]{
Comparison of $\alpha_{ii}$ and $\gamma_{i}$ calculated with the
test-field method (black solid), the reduced BS method (red dashed),
and BS method (blue dash-dotted) and plotted against
latitude in the northern hemisphere for
radii $r=0.98\,R$ (top), $r=0.84\,R$ (middle) and
$r=0.72\,R$ (bottom).
}\label{bsII}
\end{figure*}
In \cite{BS02}, a method for determining the transport
coefficients has been used which is based on the temporally varying
mean magnetic field of  
the dynamo (the main run) alone (called BS method in the following).
Instead of solving additional
test problems with predefined mean fields as described in
\Sec{sec:testfield}, the method exploits the fact that at
different times $\meanBB$ at a given position has, in general, different
directions.
So using sufficiently many time instants, the underdetermination of
\eq{eq:EMFmu} can be overcome.
One can go further and employ any available instant ending up
with a (usually heavily) overdetermined system which can be solved
approximately by the least-squares technique or singular value
decomposition.
An intrinsic problem emerges when $\meanBB$
reaches dynamically relevant strengths:
Then the transport coefficients become dependent on $\meanBB$ and
would be determined in a temporally averaged sense where, however, it
remains unclear to which strength of $\meanBB$ their values correspond.
Clearly, the BS method does not allow to obtain information on the time
evolution of the transport coefficients.
Furthermore, some of the coefficients calculated in \cite{BS02} have turned
out not to be in agreement with test-field results.
This  especially concerns the components of the turbulent diffusivity
tensor acting on currents along and perpendicular to the shear, which
are correctly obtained only with the test-field method \citep{B05AN}.

In several papers \citep{RCGBS11,SCB13,NBBMT13,ABT13,ABMT15}, the BS
method was severely simplified in that in \eq{eq:EMFmu}
the contributions with derivatives of $\meanBB$ were dropped, that is,
$\bbeta$, $\ddelta$ and $\kkappa$ were set to zero (called
reduced BS method in the following).
Such a modeling can hardly be expected to offer any predictive power
as already turbulent diffusion is not modeled in a proper way. 
Accordingly, when employing the coefficients found in this way, one
has to add turbulent diffusion by hand in order to obtain reasonable
growth rates.
With respect to descriptive power, this method must also fail 
as the individual, interpretable turbulent effects are all
subsumed in one tensor.
Recently, \cite{SCD16} used the full BS method and compared the
coefficients for one of their
simulations with the reduced BS method.
They found the $\aalpha$ and $\ggamma$ tensors to be nearly the same in
both cases and their values of $\bbeta$ to be much smaller than
their $\etatz$.

Here we demonstrate that indeed the $\aalpha$ and $\ggamma$
coefficients derived with the (reduced) BS method do not show
correspondences to those derived by the test-field method; see
\Fig{bs} and for more details \Fig{bsII}.
Comparison reveals that in many cases not even the signs are
correct, for example,\ $\alpharr^{\rm BS}$ is negative in the
northern hemisphere, close to the equator, where $\alpharr$ is
strongly positive; see \Fig{alpij}.
We also note the much smaller spatial structures in \Fig{bs} 
compared to \Figs{alpij}{turb}.
Furthermore, we also find that the reduced and the full BS give
very similar results for the $\aalpha$ and $\ggamma$ tensor.
This confirms their limited use for modeling and analysis
purposes.
It also indicates that the significantly smaller values of
$\bbeta^{\rm BS}$, compared to $\etatz$, found by \cite{RCGBS11} and
\cite{SCB13,SCD16} with the setup of \cite{GCS10}
\citep[see a comparison in Appendix~A of][]{KKOWB17},   
are not due to particularities of the latter, but
are a particular outcome of this method.
Therefore most of the conclusions derived from $\aalpha^{\rm BS}$
and $\ggamma^{\rm BS}$ in the papers cited above are to be considered misleading.
In particular, the conclusion of \cite{ABMT15} that their equatorward
migrating field is not well explained by the Parker-Yoshimura rule is
most likely unreliable.
As becomes obvious from \Fig{alp_eff}(c), using $\alphapp^{\rm BS}$
instead of $\alphapp$ in \eq{Eq:eqm} will lead to an incorrect prediction of the
propagation direction.

\section{Conclusions}

\begin{figure}[t!]
\begin{center}
\includegraphics[width=.98\columnwidth]{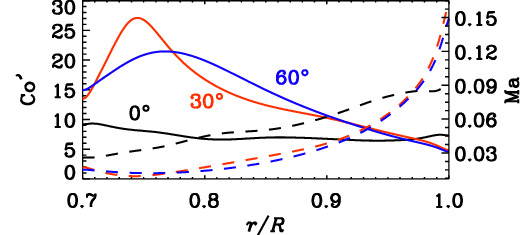}
\end{center}\caption[]{Radial dependency of Coriolis number $\Co^\prime=2\Omega_0/\urmsp \kf$
(solid lines, left $y$ axis)
and Mach number $\Ma=\urmsp/\cs$ (dashed lines, right
$y$ axis) for latitude $0^\circ$ (black), $30^\circ$ (red) and $60^\circ$ (blue).
$\cs$ is the longitudinally averaged sound speed.
}\label{cor}
\end{figure}

We have determined the full tensorial expressions of the
transport coefficients relevant for the evolution of large-scale magnetic fields
from a solar-like fully compressible global convective dynamo simulation using the
test-field method.
The simulation used here exhibits a large-scale magnetic field with solar-like
equatorward propagation similar to the runs of
\cite{KMB12,KMCWB13,KKOBWKP15,KKOWB17}.
This behavior can be well explained with the Parker-Yoshimura rule of
an $\alpha\Omega$ dynamo wave, where the negative shear at mid
latitudes generates the mean toroidal magnetic field \citep{WKKB14}.
However, we find indications that, locally,
the $\alpha$ effect is comparable to or stronger than the $\Omega$ effect
in generating toroidal magnetic field.
This suggests an $\alpha^2$ dynamo
operating locally in addition to the $\alpha\Omega$ dynamo.
This is mostly due to the dominance of $\alpharr$ among all coefficients of
$\aalpha$.
The meridional profiles of the $\alpha$ coefficients do not agree with
estimates based on helicities or simple $\cos\theta$ profiles.

The most interesting results come from the turbulent pumping velocities.
They significantly alter the effective meridional circulation and shear acting
on the mean magnetic field.
Most strikingly, the pattern of three meridional flow cells outside the tangent cylinder 
disappears completely.
Furthermore, the radial turbulent pumping at the surface is 
upward near the equator and downward at higher latitudes and therefore
in agreement with test-field results of Cartesian convection
\citep{KKB09} near the equator.
This has consequences also for possible dynamo mechanisms in the Sun.
Even if helioseismic measurements revealed a meridional
circulation pattern in favor of the flux transport dynamo \citep[see
e.g.,][]{HKC14}, the unknown turbulent pumping velocity would be able to
completely change the effective meridional circulation, in particular
as the effective velocities do not have to obey the conservation of mass.
Also the shear is altered by the presence of turbulent pumping, however
the effect is not as strong as for the meridional flow.

For the turbulent diffusion, we find that $\bbeta$ also has a small
negative definite  contribution, which can potentially lead to dynamo
action.
However, in this particular simulation, it seems to play only a minor
role.
In general, the equatorial symmetry of all time-averaged coefficients
agrees well with what is expected from equatorial symmetric flows.

If we compare the transport coefficients determined in this work with
those of previous studies of the geodynamo \citep{SRSRC07,SPD12,Schr11},
we find that most of the components agree
well in the lower part of the convection zone, but disagree    
in the upper part.
We highlight that the time-averaged $\alpharr$ and
$\alphatt$ have the opposite sign and are stronger everywhere.
We associate the agreement in the lower part of the convection zone of
our simulation with stronger rotational influence and weaker effects 
of compressibility and stratification than in the upper part and
surface region;
hence the better fit with the mentioned studies.
In \Fig{cor} we plot the radius- and latitude-dependent Coriolis and
Mach numbers;
this clearly shows the presence of two regimes: For mid and high
latitudes, the rotational influence is significantly stronger in the
lower part of the convection zone than near the surface.
In that region,
Coriolis numbers\footnote{ We use 
$\Co=l_p/\pi^2{\rm Ro}_l$ with
 $l_p$ and ${\rm Ro}_l$ from \cite{SPD12}. This gives $\Co=74, 24, 24, 20$
 for their models~4, 29, 31, and 34.} estimated from \cite{SPD12} fit
well with our values for most of their models. 
Furthermore, the Mach number decreases by about one order of magnitude
from the surface to the bottom of the convection zone; see \Fig{cor}.
This indicates that in its lower part, compressibility is less important
than near the surface.

The coefficients are altered by the presence of the magnetic field.
Comparison with a non-magnetic simulation reveals a quenching for most
of the coefficients and for the differential rotation.
However, there exist also enhancements at some localized regions for
some of the coefficients.
In the magnetic simulation itself, we found that the coefficients are
altered at the locations where the magnetic field is dynamically
important.
Most of the coefficients are quenched for strong magnetic fields, but
for example $\gamma_r$ and $\beta_{\phi\phi}$  also show enhancements,
when the field is in equipartition with the turbulent flow.
No simple algebraic expression that depends on the magnetic field strength
would be able to reproduce this kind of behavior.
Furthermore, we find a clear cyclic modulation of all coefficients, in
particular at high latitudes where the magnetic field is the
strongest.
Some of the coefficients seem to be quenched predominantly via the
toroidal magnetic field (e.g.,\ $\alphatt$, $\alphapp$) and others via
the radial field (e.g., $\alpharr$, $\gamma_r$, $\gamma_\theta$,
$\gamma_\phi$, $\bbeta$).

The overall strength of the temporal variations is
comparable with or even larger than the time-averages of the quantities,
suggesting a strong contribution to the magnetic field evolution via
these variations; they are predominantly random, but also 
show some response to the activity cycle, that is, the cyclic change in the magnetic field.
These cyclic variations are significantly stronger
than those of the mean flows indicating that the nonlinear
feedback occurs predominantly via the transport coefficients and not
via the mean flows.

The finding of \cite{WKKB14,WKKB15} that the Parker-Yoshimura rule can well
predict the propagation direction of the magnetic field can be
verified because (1) the radial $\Omega$ effect dominates the toroidal
field generation, (2) $\alphapp$ has the strongest contribution to
the radial $\alpha$ effect, and (3) $\alpha_{\rm K}$ has the same sign
as $\alphapp$ in the region of interest.
However, we find that the latitudinal derivative of $\alphapp$ also
plays a significant role.
Therefore, a detailed analysis for every simulation is necessary.

We also compare the test-field method with the 
one used by \cite{RCGBS11} and \cite{SCB13}, which is based on a
multidimensional
regression method introduced by \cite{BS02}.
Their method gives incorrect results leading to, for example, the opposite sign
of $\alpharr$ and the incorrect propagation direction from the
Parker-Yoshimura rule.

For followup work, we plan to use the determined transport
coefficients in mean-field simulations to investigate in detail their
effect and their descriptive power for the magnetic field evolution.
We also plan to investigate how the coefficients, and therefore the dynamo
mechanism, change by changing rotation, stratification and magnetic,
fluid, and entropy diffusivities.
Furthermore, it would be interesting to study the effects of a
realistic treatment of the outer magnetic boundary
\citep{WBM11,WKMB12,WKMB13,WKKB15} as well as the effect of the
spontaneous formation of magnetic flux concentrations
\citep[e.g.,][]{BKR2013,WLBKR13,WLBKR15,KBKKR16}.

\appendix

\section{Reconstruction of the electromotive force}

\begin{figure*}[t!]
\begin{center}
\includegraphics[width=\textwidth]{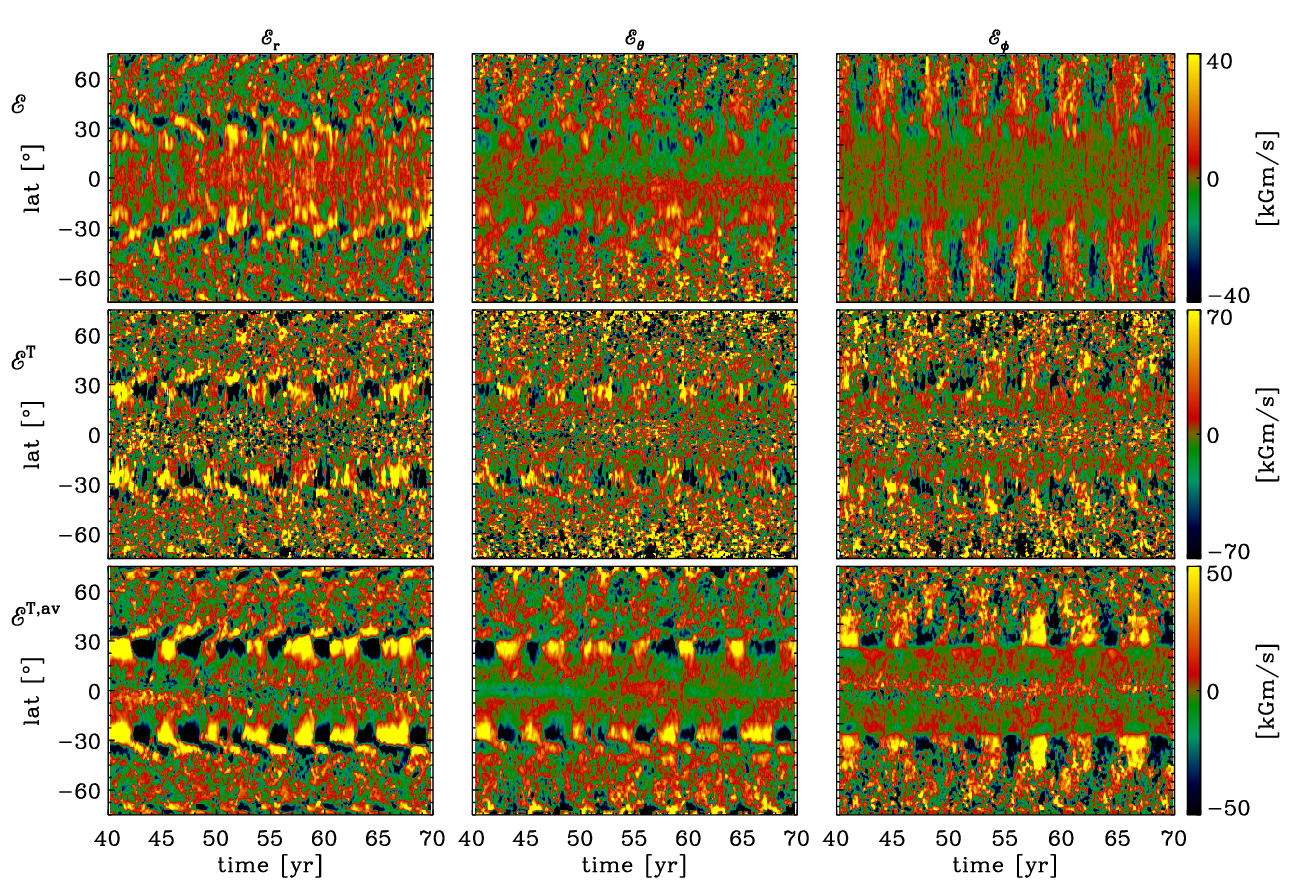}
\end{center}\caption[]{
Comparison of all components of the original electromotive force $\EMF$
(top row)
with the reconstructed 
one, $\EMF^{\rm T}$,
 in the middle of the convection
zone, $r=0.85\, R$.
Middle row:
Reconstruction with the time dependent
transport coefficients, $\EMF^{\rm T}$;
Bottom row:
reconstruction with the time-averaged    
transport coefficients, $\EMF^{\rm T,av}$, using
Eq.~\eqref{eq:emf2}. All data smoothed over 20 neighboring points in
time. 
}\label{recon}
\end{figure*}

To show that the reconstruction of the turbulent electromotive force,
employing the turbulent transport coefficients and here being labeled
$\EMF^{\rm T}$,  agrees reasonably well with the directly obtained
$\overline{\uu^{\prime}\times \bb^{\prime}}$
 (``original'' $\EMF$),  we plot in \Fig{recon} both quantities
in the middle of the convection zone.
The location of strong activity at around $\pm30^\circ$ latitudes is
well reproduced in all components. Furthermore the polarity reversals
at high latitude agree well  in the original and reconstructed $\EMF$.
Close to the equator, the agreement degrades:
In the original $\EMF$ we find a stationary positive pattern in the
radial component and an asymmetry across
the equator in the latitudinal one, but this pattern is not reproduced
in $\EMF^{\rm T}$.
Furthermore, the absolute strength of the original $\EMF$ is only nearly half 
of that of $\EMF^{\rm T}$.
If we reconstruct the electromotive force from the time-averaged
turbulent transport coefficients ($\EMF^{\rm T,av}$),
the absolute values are
closer to the original ones, but still around 30\% larger.
In all, $\EMF^{\rm T,av}$ seems to give a better reconstruction
than $\EMF^{\rm T}$, but the large random variations in time
might blur the comparison (see \Sec{sec:var}).
We associate the differences between reconstructed and original $\EMF$
with lack of scale separation, that is, non-locality in space and time
\citep{BRS08,HB09,RB12}
and will address these issues in forthcoming publications.

\section{Comparison with mean-field model}
\label{sec:appB}

\begin{figure*}[t!]
\begin{center}
\includegraphics[width=0.8\textwidth]{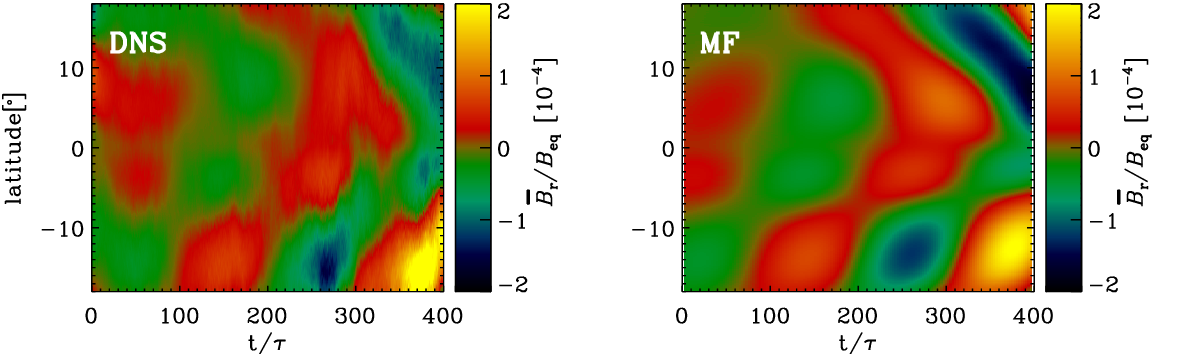}
\end{center}\caption[]{
$\mean{B}_r$ from a DNS and a corresponding mean-field model employing
the turbulent transport coefficients, obtained with the test-field
method from the same DNS, as functions of latitude and time at $r=0.94\,R$, where
$\tau=1/\urms\kf$ is the turbulent turnover time. 
}\label{mf_dns}
\end{figure*}

To show how well the transport coefficients describe and predict the
mean magnetic field, we performed additional
direct numerical simulations
(DNS) producing an oscillating
spherical dynamo from 
forced turbulence.
The setup is similar to \cite{MTKB10}, but 
with forcing wavenumber $\kf/k_1=10$, where $k_1=2\pi/0.3R$
corresponds to the shell thickness, and $\Rey=0.63$, $\Rm=0.96$.
We computed the turbulent transport coefficients using the 
presented method and solved a corresponding mean-field model,
which reproduces three key features of the field
evolution in the DNS: (1) The growth rate of the magnetic field
($\lambda_{\rm DNS}=0.00558/\tau$, $\lambda_{\rm MF}=0.00570   
/\tau$ for the volume averaged rms field),
(2) its oscillation period 
($T_{\rm DNS}=109\, \tau$, $T_{\rm MF}=113\, \tau$)
and (3) its complex latitudinal distribution.
\Fig{mf_dns} shows the radial
mean magnetic field as a function of time and latitude from both the DNS and the mean-field model.
It is rather expected that the correspondence is very good in this case,
as the scale separation is high and the Reynolds numbers are small.
In our compressible convection simulations these conditions are no
longer fulfilled, hence non-local effects in space and time likely
play an important role. Therefore, comparison to mean-field models is
less trivial and will be addressed in detail
in a forthcoming publication.

\section{Comparison with analytical results}
\label{sec:appC}
\begin{figure}[t!]
\begin{center}
\includegraphics[width=\linewidth]{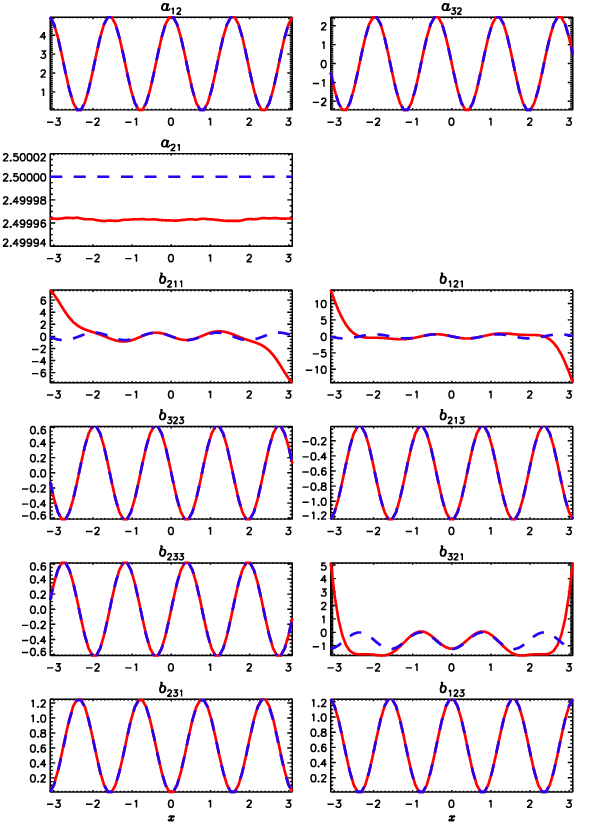}
\end{center}\caption[]{Comparison of analytical and test-field results for the flow
\eq{eq:rob} showing the mean-field coefficients as a
function of $x$.
Red solid: numerical, blue dashed: analytical.
$v_0=w_0=1$, $k_0=2$, $\eta=0.05$ in arbitrary units.
The discrepancies in $b_{211}, b_{121}$, and $b_{321}$ near the $x$
boundaries are systematic errors due to the use of periodic boundary
conditions for the solution of \eq{eq:testpr}, which conflicts with
the test fields, 
being
 linear in $x$ and $z$.
}\label{fig:coeffs}
\end{figure}

For comparison with analytic results we have chosen the flow
\begin{equation} 
\uu = 
\begin{pmatrix}
~~v_0 \sin k_0 x \cos k_0 y\cr
 -v_0 \cos k_0 x \sin k_0 y\cr
~~w_0 \cos k_0 x \cos k_0 y
\end{pmatrix},
\label{eq:rob}
\end{equation}
in a Cartesian ($x,y,z$) domain,
out of a family of three, introduced by \cite{Rob70};
see \cite{Rhetal14} for comparison.
Here, $v_0$ and $w_0$ are constant prefactors and
$k_0$ is the horizontal wavenumber of the flow.
Under SOCA and for $\eta=\const$, we obtain, with averaging over $y$,
for the coefficients of \eq{eq:emf1},
\begin{align}
  a_{11} &=  a_{31}  =  a_{22} =  a_{i3} =0, \quad i=1,2,3, \\
 a_{12} &=   \frac{v_0 w_0}{2 \eta k_0}  \cos^2 k_0 x, \quad a_{21} =  \frac{v_0 w_0}{4 \eta k_0}, \\
   a_{32} &=   - \frac{v_0^2}{2 \eta k_0}  \sin k_0 x \cos k_0 x, \\
   b_{11i} &= b_{31i} = b_{22i} = b_{13i} = b_{33i} =0, \quad i=1,3, \\
    b_{211} &= b_{121} = b_{323} = - b_{233} = -\frac{v_0 w_0}{4 \eta k_0^2}  \sin k_0 x \cos k_0 x, \\
   b_{321} &= -\frac{v_0^2}{4 \eta k_0^2}  \cos^2 k_0 x, \quad
   b_{231} = \frac{v_0^2}{4 \eta k_0^2}  \sin^2 k_0 x, \\
   b_{213} &=  -\frac{w_0^2}{4 \eta k_0^2}  \cos^2 k_0 x = -b_{123}.
  \end{align}
  Figure~\ref{fig:coeffs} shows these profiles in comparison with the results of the test-field method applied in Cartesian geometry. 
  To enable maximal agreement, the non-SOCA term 
  $\uu^{\prime}\times\bb^{\prime}-\overline{\uu^{\prime}\times\bb^{\prime}}$  of \eq{eq:testpr}
   was switched off in the code.

\begin{acknowledgements}
The authors thank the referee Ludovic Petitdemange for numerous useful
comments and suggestions, which improved the paper.
The simulations have been carried out on supercomputers at
GWDG, on the Max Planck supercomputer at RZG in Garching, in the HLRS
High Performance Computing Center Stuttgart, Germany through the PRACE
allocation ``SOLDYN'' and in the facilities hosted by the CSC---IT
Center for Science in Espoo, Finland, which are financed by the
Finnish ministry of education. 
J.W.\ acknowledges funding by the Max-Planck/Princeton Center for
Plasma Physics and 
 from the People Programme (Marie Curie
Actions) of the European Union's Seventh Framework Programme
(FP7/2007-2013) under REA grant agreement No.\ 623609.
This work was partially supported by the 
University of Colorado through its support of the
George Ellery Hale visiting faculty appointment, the
Swedish Research Council grants No.\ 621-2011-5076 and 2012-5797 (A.B.),
and the Academy of Finland Centre of Excellence ReSoLVE 272157
(J.W., M.R., S.T., P.J.K., M.J.K.) and grants 136189, 140970, 272786 (P.J.K).

\end{acknowledgements}

\bibliographystyle{aa}
\bibliography{paper}

\end{document}